\UseRawInputEncoding
\documentclass[11pt,a4paper]{article}
\pdfoutput=1

\usepackage{jcappub}
\usepackage{amsmath}
\usepackage{amssymb}
\usepackage{natbib}
\usepackage{graphicx}
\usepackage{lscape}
\usepackage{hyperref}
\usepackage{xcolor}
\usepackage[normalem]{ulem}

\newcommand{\Lya}{Ly$\alpha$}
\newcommand{\Lyb}{Ly$\beta$}

\newcommand{\angstrom}{\mbox{\normalfont\AA}}

\newcommand{\secref}[1]{\S~\ref{#1}}
\newcommand{\phm}{\phantom{$-$}}
\newcommand{\phc}{\phantom{$0$}}

\title{\texttt{LyaCoLoRe}: Synthetic Datasets for Current and Future Lyman-$\alpha$ Forest BAO Surveys}

\author[a]{James Farr \href{https://orcid.org/0000-0002-9817-533X}{\includegraphics[scale=0.096]{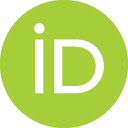}},}

\author[a]{Andreu Font-Ribera,}
\author[b]{H\'{e}lion du Mas des Bourboux,}
\author[c]{Andrea Mu\~{n}oz-Guti\'{e}rrez,}
\author[d,e]{F. Javier S\'{a}nchez \href{https://orcid.org/0000-0003-3136-9532}{\includegraphics[scale=0.096]{ORCIDiD_icon128x128.png}},}
\author[a]{Andrew Pontzen,}
\author[f,g]{Alma Xochitl Gonz\'{a}lez-Morales,}

\author[h]{David Alonso,}
\author[a]{David Brooks,}
\author[a]{Peter Doel,}
\author[i]{Thomas Etourneau,}
\author[j]{Julien Guy,}
\author[i]{Jean-Marc Le Goff,}
\author[c]{Axel de la Macorra,}
\author[i]{Nathalie Palanque-Delabrouille,}
\author[k]{Ignasi P\'{e}rez-R\`{a}fols,}
\author[i]{James Rich,}
\author[l]{An\v{z}e Slosar,}
\author[m]{Gregory Tarle,}
\author[n]{Duan Yutong,}
\author[j]{and Kai Zhang}

\affiliation[a]{University College London, Gower Street, London, WC1E 6BT, UK}

\affiliation[b]{Department of Physics and Astronomy, University of Utah, 115 S. 1400 E., Salt Lake City, UT 84112, USA}

\affiliation[c]{Instituto de F\'{i}sica, Universidad Nacional Aut\'{o}noma de M\'{e}xico, A.P. 70-264, 04510, M\'{e}xico D.F., M\'{e}xico}

\affiliation[d]{Department of Physics and Astronomy, University of California, Irvine, CA 92697, USA}

\affiliation[e]{Fermi National Accelerator Laboratory, P.O. Box 500, Batavia, IL, USA}

\affiliation[f]{Departamento de F\'{i}sica, DCI, Campus L\'{e}on, Universidad de Guanajuato, 37150, L\'{e}on, Guanajuato, M\'{e}xico}

\affiliation[g]{Consejo Nacional de Ciencia y Tecnolog\'{i}a, Av. Insurgentes Sur 1582. Colonia Cr\'{e}dito Constructor, Del. Benito Ju\'{a}rez, C.P. 03940, M\'{e}xico D.F., M\'{e}xico}

\affiliation[h]{Department of Physics, University of Oxford, Keble Road, Oxford, OX1 3RH, UK}

\affiliation[i]{IRFU, CEA, Universit\'{e} Paris-Saclay, F-91191 Gif-sur-Yvette, France}

\affiliation[j]{Lawrence Berkeley National Laboratory, 1 Cyclotron Road, Berkeley, CA 94720, USA}

\affiliation[k]{Sorbonne Universit\'{e}, CNRS/IN2P3, Laboratoire de Physique Nucl\'{e}aire et de Hautes Energies, LPNHE, 4 Place Jussieu, F-75252 Paris, France}

\affiliation[l]{Brookhaven National Laboratory, 2 Center Road, Upton, NY 11973, USA}

\affiliation[m]{Physics Department, University of Michigan Ann Arbor, MI 48109, USA}

\affiliation[n]{Physics Department, Boston University, 590 Commonwealth Avenue, Boston, MA 02215, USA}

\emailAdd{james.farr.17@ucl.ac.uk}

\abstract{The statistical power of Lyman-$\alpha$ forest Baryon Acoustic Oscillation (BAO) measurements is set to increase significantly in the coming years as new instruments such as the Dark Energy Spectroscopic Instrument deliver progressively more constraining data. Generating mock datasets for such measurements will be important for validating analysis pipelines and evaluating the effects of systematics. With such studies in mind, we present \texttt{LyaCoLoRe}: a package for producing synthetic Lyman-$\alpha$ forest survey datasets for BAO analyses. \texttt{LyaCoLoRe} transforms initial Gaussian random field skewers into skewers of transmitted flux fraction via a number of fast approximations. In this work we explain the methods of producing mock datasets used in \texttt{LyaCoLoRe}, and then measure correlation functions on a suite of realisations of such data. We demonstrate that we are able to recover the correct BAO signal, as well as large-scale bias parameters similar to literature values. Finally, we briefly describe methods to add further astrophysical effects to our skewers --- high column density systems and metal absorbers --- which act as potential complications for BAO analyses.}

\arxivnumber{1912.02763}
\begin{document}
\maketitle

\section{Introduction}

Our understanding of the expansion history of the Universe has progressed enormously over the last quarter of a century. The discovery of accelerating expansion from the ``standardisable candles'' of supernovae~\cite{Riess:1998AJ....116.1009R,Perlmutter:1999ApJ...517..565P} brought the idea of dark energy to the fore, and it is now considered a vital component of the cosmic inventory. Indeed, efforts to improve our measurements of its properties are at the forefront of current cosmological research, and it is a primary motivation behind a number of surveys past, present and future.

Several of these surveys have focussed on using the ``standard ruler'' of Baryon Acoustic Oscillations (BAO)~\cite{Peebles:1970ApJ...162..815P} in their efforts to understand the Universe's expansion. This fixed-scale imprint on structure formation was first measured from the correlation function~\cite{Eisenstein:2005ApJ...633..560E} and power spectrum~\cite{Cole:2005MNRAS.362..505C} of galaxy samples from the Sloan Digital Sky Survey (SDSS) and 2dF Galaxy Redshift Survey respectively. A number of similar measurements have been made in subsequent years, focussing on using galaxies~\cite[e.g.][]{Percival:2010MNRAS.401.2148P,Beutler:2011MNRAS.416.3017B,Blake:2011MNRAS.418.1707B,Alam:2017MNRAS.470.2617A} and quasars (QSOs)~\cite[e.g.][]{Ata:2018MNRAS.473.4773A} as tracers of the matter density. These tracers cover redshift ranges $z\sim 0.1-1.0$ and $z\sim 1.2-1.7$ respectively.

An alternative tracer exists in the form of the Lyman-$\alpha$ (\Lya) forest: a sequence of absorption features that appears in the spectra of high-$z$ QSOs as a result of \Lya\ absorption of light in the neutral hydrogen gas between QSO and observer. These spectral features thus trace the density of neutral hydrogen gas in the inter-galactic medium (IGM) along the line of sight \cite{Bi:1992A&A...266....1B}. Indeed, analytical models developed during the 1990s showed that the \Lya\ forest absorption closely traces the distribution of dark matter on scales larger than the Jeans length \cite[e.g.][]{Cen:1994ApJ...437L...9C,Petitjean:1995A&A...295L...9P,Miralda-Escudé:1996ApJ...471..582M}. The \Lya\ forest should, then, provide a suitable means to extend measurements of cosmic expansion via BAO to earlier in the Universe's history. Measuring such a signal was first discussed in~\cite{McDonald:2007PhRvD..76f3009M}, while the 3D correlation of flux transmission was first studied in~\cite{Slosar:2011JCAP...09..001S}. The BAO signal was first detected from measurements of the \Lya\ auto-correlation using data from data release 9 (DR9) of the Baryon Oscillation Spectroscopic Survey (BOSS) of SDSS-III~\cite{Busca:2013A&A...552A..96B,Slosar:2013JCAP...04..026S,Kirkby:2013JCAP...03..024K}, with subsequent improvements in DR11~\cite{Delubac:2015A&A...574A..59D} and DR12~\cite{Bautista:2017A&A...603A..12B}, as well as DR14 of the extended Baryon Oscillation Spectroscopic Survey (eBOSS)~\cite{deSainteAgathe:2019A&A...629A..85D}. The cross-correlation between the \Lya\ forest and QSOs was first measured in BOSS DR9~\cite{Font-Ribera:2013JCAP...05..018F}, with the first detection of BAO coming in DR11~\cite{Font-Ribera:2014JCAP...05..027F}, and improvements made in DR12~\cite{duMasdesBourboux:2017A&A...608A.130D} and eBOSS DR14~\cite{Blomqvist:2019A&A...629A..86B}.

The upcoming Dark Energy Spectroscopic Instrument (DESI)~\cite{DESICollaboration:2016arXiv161100036D} will be able to advance these measurements greatly. Over the 5 years of its operation, it will measure approximately 800,000 QSO spectra with $z>2.0$, 3 times as many as in the final eBOSS dataset (approximately 270,000). Ahead of such an increase in statistical power, it is vital to be able to sufficiently test analysis pipelines to ensure that they do not introduce any biases. Equally, it is important to be able to quantify exactly how secondary astrophysical effects will impact upon BAO measurements. The best way to carry out both of these tests is through the development of mock datasets \cite[e.g.][]{LeGoff:2011A&A...534A.135L,Font-Ribera:2012JCAP...01..001F,Bautista:2015JCAP...05..060B} --- synthetic realisations of a survey for which cosmological and astrophysical parameters can be easily controlled. Producing such datasets must be computationally inexpensive in order to allow for generation of a large number of realisations, but the data must also provide realistic representations of the survey itself.

In this work, we introduce a package designed to produce mock datasets for current and future \Lya\ forest BAO analyses, \texttt{LyaCoLoRe}. In \secref{sec:Making the mocks}, we describe the methods used to generate such datasets, including the use of a Gaussian random field to generate the 3D correlations and the subsequent post-processing to yield realistic skewers of transmitted flux fraction. The methods to determine the optimal values of parameters used in these transformations are detailed in \secref{sec:Parameter tuning}. We then verify that the datasets are able to fulfil their purpose for BAO analyses in \secref{sec:Verifying the mocks}, measuring correlation functions in the same way as recent analyses from BOSS and eBOSS. In \secref{sec:Adding secondary astrophysical effects}, we introduce and briefly test additional astrophysical effects that \texttt{LyaCoLoRe} is able to include, before summarising and concluding in \secref{sec:Summary conclusions}.

\section{Making the mocks}
\label{sec:Making the mocks}

The requirement of mocks to be computationally inexpensive but also large in volume prohibits the use of hydrodynamical or full N-body simulations in their construction. Instead, Gaussian random field methods can be used to generate a linear density field in a large box. This method does not capture non-linear evolution, generating data based solely on an initial power spectrum, but is orders of magnitude faster than state of the art simulations. From an initial Gaussian field, a number of options are available to model the physical density. Most straightforwardly, using a lognormal approximation provides a semi-analytic and physically plausible ($\rho>0$) model of the density field, but breaks down beyond weakly non-linear scales \cite{Coles:1991MNRAS.248....1C,Bi:1997ApJ...479..523B}. Formalisms such as Lagrangian perturbation theory \cite[see][for a review]{Bernardeau:2002PhR...367....1B} can help extend to mildly non-linear scales, while COLA \cite{Tassev:2013JCAP...06..036T} methods subsequently use a small number of N-body code timesteps to further improve modelling of non-linearities. The choice of density approximation method depends on the scales of interest and the computational budget for the task at hand. The presence of non-linear structure is not of vital importance to BAO measurements, particularly at $z\geqslant2$ where the \Lya\ forest is observed \cite{Kirkby:2013JCAP...03..024K}. As such, Gaussian random field methods are well suited to the production of \Lya\ BAO mock datasets, and using a lognormal approximation for the physical density is sufficient for current analyses \cite{LeGoff:2011A&A...534A.135L}. Studies that are more heavily dependent on accurate reproduction of small-scale effects may require alternative techniques to be used. Having generated a physical density field, tracers such as QSOs can be placed at its peaks via Poisson sampling according to an input bias and number density, and line-of-sight skewers can be drawn by interpolating within the box.

Converting density skewers to mimic the transmitted flux fraction of the \Lya\ forest then requires a significant degree of post-processing. Despite the speed of Gaussian random field methods, resolution higher than $O(1)$~Mpc/$h$ is not possible within the computational bounds of mock production due to memory limitations. As a result, the 1D power spectrum of the skewers $P_{1\mathrm{D}}(k_\parallel)$ --- the power spectrum measured only from modes lying along the line of sight of each skewer --- is greatly suppressed. This subsequently affects the errors on our BAO measurements, as the 3D flux power spectrum of the \Lya\ forest has a significant contribution to its error that is proportional to the 1D power spectrum, known as aliasing noise~\cite{McDonald:2007PhRvD..76f3009M}. As such, we must boost the 1D power spectrum by the addition of small-scale fluctuations in order to ensure that our BAO errors behave correctly. Further, we must convert from density to optical depth at each point of each skewer. The details of this relationship are complex, but in the context of Gaussian random field mocks we are constrained to using a simple approximation such as the fluctuating Gunn-Peterson approximation (FGPA) \cite{Croft:1998ApJ...495...44C}. Methods such as those discussed in \cite{Irsic:2018JCAP...04..026I} offer improved physical intuition behind the structure of the \Lya\ forest, but developing techniques to apply them in the context of full-survey mock datasets is beyond the scope of this work. Finally, we must add redshift-space distortions to our skewers. These distortions occur as a result of peculiar velocities in the IGM, and we observe them as an anisotropy in measurements of power spectra and correlation functions.

In this work, we use \texttt{CoLoRe}~\cite{Alonso:2020inprep} to generate our initial Gaussian skewers, as described in \secref{subsec:CoLoRe}. We then present the package \texttt{LyaCoLoRe}, which is able to convert \texttt{CoLoRe}'s output into realistic skewers of transmitted flux fraction. The methods used in this transformation are described in \secref{subsec:LyaCoLoRe}. Finally, in \secref{subsec:Computational requirements}, we discuss the computational requirements of running both of these packages. The output skewers from \texttt{LyaCoLoRe} then require the addition of instrumental noise and combination with a QSO continuum before they can be considered realistic spectra. This can be carried out in the context of DESI by a package called \texttt{desisim}\footnote{Publicly available at \href{https://github.com/desihub/desisim}{https://github.com/desihub/desisim}.}, which is not discussed in this work.

\subsection{\texttt{CoLoRe}: Cosmological Lognormal Realisations}
\label{subsec:CoLoRe}

The \texttt{LyaCoLoRe} mocks originate from a program called \texttt{CoLoRe}\footnote{Publicly available at \href{https://github.com/damonge/CoLoRe}{https://github.com/damonge/CoLoRe}.}, a highly parallelised code initially designed to produce large catalogues of multiple tracers with the same underlying density field \cite{Alonso:2020inprep}. In this work, we use \texttt{CoLoRe}'s lognormal density model for speed, though first and second order Lagrangian perturbation theory methods are also available. Making use of these functionalities would constitute a natural extension of this work, and efforts are ongoing to do so.

From this density field, \texttt{CoLoRe} can produce a number of observables such as cosmic shear, intensity maps, CMB lensing and integrated Sachs-Wolfe maps. Most importantly in the context of this work, it is also able to draw line-of-sight skewers from each object to a central observer, interpolating the Gaussian field at intermediate points. This final functionality makes \texttt{CoLoRe} well suited for \Lya\ forest mocks. The basic steps that \texttt{CoLoRe} takes in computing such skewers are outlined in the 5-stage process below:

\begin{enumerate}
    \item Generate a Gaussian random field $\delta_C$ at $z=0$ in a Cartesian box according to an input power spectrum.
    \item Compute a corresponding radial velocity in each cell using the gradient of the Newtonian gravitational potential $\phi$:
     \begin{equation}
        v_r(z=0) = -\frac{2f_0}{3 H_0^2 \Omega_M} (\mathbf{e}_r \cdot \nabla) \phi(z=0),
    \end{equation}
    where $f_0$ is the logarithmic growth rate at $z=0$, $H_0$ is the Hubble constant, $\Omega_M$ is the matter density parameter, and $\mathbf{e}_r$ is the radial unit vector.
    \item Calculate the redshift of each cell (taking the centre of the box as the observer) using a given input cosmology, and de-evolve the fields to that redshift using the corresponding linear growth factor.
    \item Carry out a lognormal transformation of the Gaussian field, and Poisson sample it using an input number density $n(z)$ and bias $b(z)$ to obtain a set of sources (QSOs in our case).
    \item Compute line-of-sight skewers from each source to the centre of the box by interpolating the initial Gaussian field and the radial velocity field.
\end{enumerate}

The final output from \texttt{CoLoRe} is a set of QSOs and corresponding Gaussian field skewers, as well as values of cosmological variables along the skewers. The QSOs have the correct 3D clustering properties on large scales, as demonstrated briefly in Appendix~\ref{sec:Quasar auto-correlation} and in more detail in \cite{Alonso:2020inprep}. The skewers also have the correct 3D correlations, as demonstrated in \secref{sec:Verifying the mocks}.

\subsection{\texttt{LyaCoLoRe}}
\label{subsec:LyaCoLoRe}
\begin{figure}
\centering
\includegraphics[height=0.8\textheight]{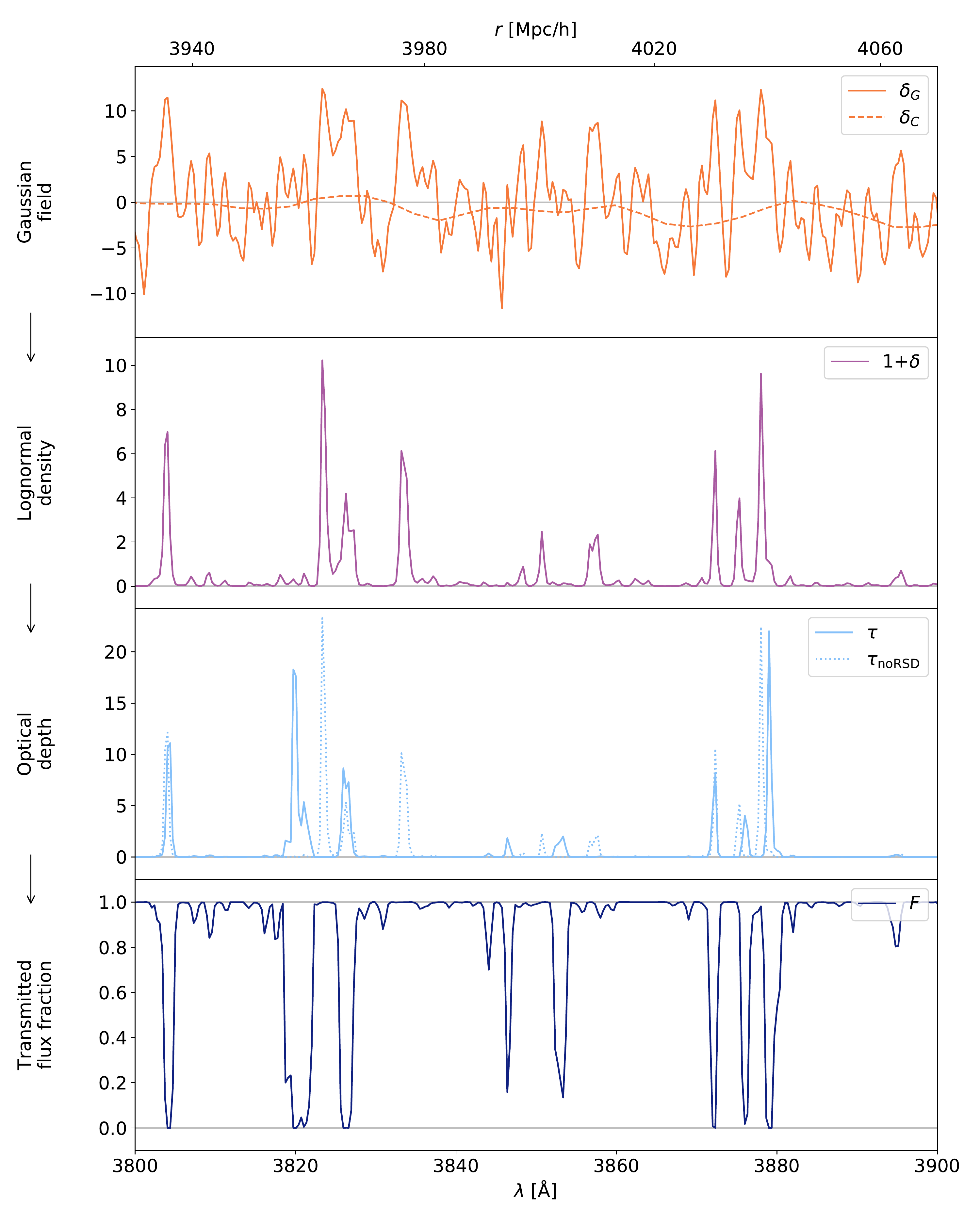}
\caption{A sample skewer shown at the different stages of transformation from ``raw'' Gaussian \texttt{CoLoRe} output to a final \texttt{LyaCoLoRe} flux skewer. The top panel shows the addition of small-scale power to the skewer as described in \secref{subsubsec:Adding small-scale power}, converting $\delta_C$ to $\delta_G$. The transition to the second panel shows the lognormal transformation from \secref{subsubsec:Transformation to optical depth}, and moving to the dotted line of the third panel shows the fluctuating Gunn-Peterson approximation (FGPA) transformation from the same section. The application of redshift-space distortions (RSDs), as described in \secref{subsubsec:Adding redshift-space distortions}, shifts the dotted line to the solid line in this third panel. The final transformation from optical depth to flux, as described in \secref{subsubsec:Final transmission skewers}, maps the third to the bottom panel. Here, the Hubble flow is used to map distances (top horizontal axis) to observed wavelengths (bottom horizontal axis).}
\label{fig:skewers}
\end{figure}

While \texttt{CoLoRe} is able to produce skewers with 3D, large-scale correlations matching a given input in a short timeframe, its ``raw'' output requires significant post-processing before it can be considered a realistic representation of the \Lya\ forest. To implement these stages of processing, we have developed a \texttt{Python} module under the name \texttt{LyaCoLoRe}\footnote{Publicly available at \href{https://github.com/igmhub/LyaCoLoRe}{https://github.com/igmhub/LyaCoLoRe}.}. This code transforms \texttt{CoLoRe}'s output into realistic skewers of transmitted flux fraction. The following sections describe the key methods that \texttt{LyaCoLoRe} uses to do so, with each step represented visually in Figure \ref{fig:skewers}.

\subsubsection{Adding small-scale power}
\label{subsubsec:Adding small-scale power}
In order that the memory requirements of running \texttt{CoLoRe} do not become overwhelmingly large, we are limited to using a grid of $4096^3$ cells. Requiring that this encloses the volume of a full \Lya\ survey limits us to using a low-resolution grid, with cells in \texttt{CoLoRe}'s raw output of $O(1)$~Mpc/$h$. In the context of the \Lya\ forest, we observe clustering on scales down to the Jeans Length, approximately 100 kpc/$h$ \cite{Walther:2018ApJ...852...22W} and an order of magnitude lower than the resolution we can feasibly achieve. While BAO is a large-scale phenomenon, imposing that the synthetic data has approximately the right small-scale properties ensures that the covariance matrices in our final analyses are realistic. We address this by first interpolating \texttt{CoLoRe}'s Gaussian skewers --- labelled as $\delta_C$ --- to a smaller cell size, using nearest grid point (NGP) interpolation in order to avoid introducing additional smoothing.

We then generate a set of new, independent Gaussian skewers $\delta_\epsilon$ on the grid of smaller cells according to an input 1D power spectrum. We take the $k$-dependence of this 1D power spectrum to follow that used in \cite{McDonald:2006ApJS..163...80M}:
\begin{equation} \label{extra power}
    P_{\rm{1D}}(k)~\propto~[1~+~(k/k_1)^n]^{-1},
\end{equation}
where the normalisation is chosen to ensure unit variance. The additional skewers are then scaled by a common factor in order to control the variance in the extra power added. This factor is allowed to vary along the length of the skewers, effectively adding a redshift-dependency to the extra power. Hence, we write this factor as $\sigma_{\epsilon}(z)$. The parameters $n$ and $k_1$, as well as the function $\sigma_{\epsilon}(z)$ are free, and we choose them according to the process described in \secref{sec:Parameter tuning}, aiming to achieve the correct 1D power spectrum across a range of redshifts. The new skewers are then simply added to each of the existing ones to form our final Gaussian skewers $\delta_G$:
\begin{equation} \label{gaussian}
    \delta_G(z,\textbf{x}) = \delta_C(\textbf{x}) + \sigma_\epsilon(z) \delta_\epsilon(\textbf{x}).
\end{equation}
The top panel of Figure \ref{fig:skewers} shows a sample skewer before and after the extra small-scale power is added. As the additional skewers are independent from one another, there are no correlations between the structures added to each of the skewers. When we measure the 3D correlation function, we ignore contributions from pixel-pairs in the same skewer and so this process of adding small-scale power will not affect the 3D correlations of the Gaussian field beyond simply adding noise.

It is worth noting that we could have chosen to add extra small-scale fluctuations to the velocity field and achieved the same correct 1D power spectrum. However, allowing parameters describing extra small-scale velocities to vary freely would require the re-computation of the redshift-space distortions weights matrix (see \secref{subsubsec:Adding redshift-space distortions}) at each step of the tuning process (see \secref{sec:Parameter tuning}). This is a considerably more time-consuming procedure than simply carrying out the inverse Fourier transform of equation \eqref{extra power}. As such, we choose to only add small-scale fluctuations to the Gaussian field and assign to each of the small cells the velocity of the nearest large \texttt{CoLoRe} cell.

\subsubsection{Transformation to optical depth}
\label{subsubsec:Transformation to optical depth}

In \texttt{LyaCoLoRe}, the transformation from skewers of the Gaussian field to ones of optical depth is governed by two equations. The first of these is known as a lognormal transformation. This approximates the density of the baryonic matter field closely by using a lognormally-distributed variable~\cite{Bi:1997ApJ...479..523B}, introducing a degree of non-linearity. This is normalised so that we may define a deviation $\delta$ from the mean density as:
\begin{equation} \label{lognormal}
    1 + \delta(z,\textbf{x}) = \frac{\rho(z,\textbf{x})}{\overline{\rho}(z)} = \exp{} \Bigg[ D(z)\delta_G(z,\textbf{x}) - D^2(z)\frac{\sigma_G^2(z)}{2} \Bigg],
\end{equation}
where $D(z)$ is the linear growth factor at redshift $z$; $\delta_G(z,\textbf{x})$ is the Gaussian field value from equation \eqref{gaussian}; $\sigma_G(z)$ is the standard deviation of this Gaussian field and $\rho(z,\textbf{x})$ is the lognormal density at redshift $z$ and position $\textbf{x}$. This transformation is shown by the transition from the top to the second panel in Figure \ref{fig:skewers}.

The second equation allows us to transform these deviations in density into an approximation of the optical depth at each point. Assuming adiabatic expansion implies a tight relationship between temperature and density of the form $\mathrm{d}\ln T/\mathrm{d}\ln \rho = \gamma-1$~\cite{Hui:1997MNRAS.292...27H}. If we further assume photoionization equilibrium, the temperature of the gas approximately determines the number of neutral hydrogen atoms $n_\mathrm{HI} \propto \rho^2 T^{-0.7}$ for a given baryonic matter density $\rho$~\cite{Hui:1997ApJ...486..599H}. As the optical depth $\tau$ is proportional to $n_{\mathrm{HI}}$~\cite{Gunn:1965ApJ...142.1633G}, these two assumptions allow us to provide an approximation for $\tau$ given $\rho$ known as the fluctuating Gunn-Peterson approximation (FGPA)~\cite{Bi:1997ApJ...479..523B,Croft:1998ApJ...495...44C}:
\begin{equation} \label{FGPA}
    \tau(z,\textbf{x}) = \tau_0(z) [1+\delta(z,\textbf{x})] ^{\alpha(z)},
\end{equation}
where $\tau_0(z)$ is a normalisation determined by the gas temperature and the photoionisation rate, and $\alpha(z)=2-0.7(\gamma(z)-1)$ is determined by the temperature-density relation. These parameter functions $\tau_0(z)$ and $\alpha(z)$ are free, and the method for choosing them is described in \secref{sec:Parameter tuning}. The transformation to optical depth is shown by the transition from the second panel to the dotted line of the third panel in Figure \ref{fig:skewers}.

\subsubsection{Adding redshift-space distortions}
\label{subsubsec:Adding redshift-space distortions}

The \Lya\ forest exists as a sequence of absorption features due to the gradient in the recessional velocity of the IGM caused by the Universe's expansion. Features are redshifted according to their distance from the observer, appearing in a spectrum at an observed wavelength $\lambda_\mathrm{obs}=\lambda_\alpha(1+z)$ for $\lambda_\alpha$ the \Lya\ wavelength, and $z$ the absorption redshift. However, peculiar velocities in a region of gas cause its redshift to differ from that due to expansion alone. These effects are known as redshift-space distortions (RSDs), and can be induced by a number of different effects. In particular, RSDs due to gravitationally-induced linear velocities in the IGM are calculated by \texttt{CoLoRe}: as mentioned in \secref{subsec:CoLoRe}, it produces velocity skewers quantifying this effect by calculating the gradient of the Newtonian gravitational potential.

The transition from real- to redshift-space in each skewer can be thought of as an integral over velocity space of the real-space optical depth field multiplied by a kernel $K$:
\begin{equation}
\label{RSD convolution}
    \tau({s}) = \int \tau({x}) K\Big({s}-{x}-v_r\big({x}| T({x})\big)\Big) \mathrm{d}{x},
\end{equation}
where $x$ and $s$ are velocity coordinates along the skewer in real- and redshift-space respectively, $v_r$ is the radial peculiar velocity, and $T$ is the temperature. The choice of $K$ depends on the complexity of the physical effects that you wish to capture. Choosing a suitable Gaussian kernel allows the inclusion of thermal broadening effects: the apparent spreading of the gas's optical depth contribution in redshift-space due to random thermal velocities of the gas atoms. This is implemented as an option within \texttt{LyaCoLoRe}, the details of which are described in Appendix \ref{sec:RSD implementation}. However, we find that the width $\sigma_v$ of this Gaussian kernel is often smaller than the typical cell size used in \texttt{LyaCoLoRe} when adding small-scale fluctuations. Thus, the net effect of accounting for this physical process is small, and so for the purposes of this work we choose the most straightforward option, setting $K(x) = \delta^D(x)$ for $\delta^D$ the Dirac delta function. This shifts the optical depth along each skewer according to the peculiar velocity, and does not attempt to include any further physical effects.

In order to implement equation \eqref{RSD convolution}, we determine a matrix of weights $W_{ij}$ for each skewer to map its real-space cells $\tau^x_j$ to redshift-space cells $\tau^s_i$ via the matrix equation $\tau^s_i~=~W_{ij}\tau^x_j$. The matrix $W_{ij}$ depends on the velocities in the skewer as well as the choice of kernel $K$, and the details of its calculation can be found in Appendix~\ref{sec:RSD implementation}. Our implementation conserves the integrated optical depth along each line of sight (ignoring pixels which are shifted to un-observed wavelengths). The matrix $W_{ij}$ is near-diagonal and filled mostly by zeros. It can thus be stored in the form of a sparse matrix, and applied to any additional absorption transitions (see \secref{subsec:Including additional absorption transitions}), reducing both the computation time and memory requirements of adding RSDs to the skewers.

The addition of RSDs (without thermal broadening) to a sample optical depth skewer is shown by the transition from the dotted to the solid line in the third panel of Figure~\ref{fig:skewers}.

\subsubsection{Final transmission skewers}
\label{subsubsec:Final transmission skewers}

In one final stage, we convert from skewers of optical depth $\tau$ to transmitted flux fraction $F$ via the equation:
\begin{equation} \label{tau_to_flux}
   F(s) = \exp{} \big[ -\tau(s) \big],
\end{equation}
and interpolate onto a wavelength grid of the user's choice to obtain $F(\lambda)$, where $\lambda = \lambda_\alpha (1+z)$. These skewers are then written to disc.

This final transformation can be seen in the transition between the solid lines in the third and fourth panels of Figure \ref{fig:skewers}. It is worth noting that, while the signal in the lognormal density deviation $1+\delta$ and optical depth $\tau$ skewers is dominated by over-dense regions, the signal in flux $F$ becomes saturated (equal to 0) at these points and does not carry a great deal of information. Rather, the intermediate density regions --- where the density is high enough to cause some absorption but not so high that saturation occurs --- are those from which the most information can be gleaned.

\subsection{Computational requirements}
\label{subsec:Computational requirements}

In the realisations presented in this work, we specify that \texttt{CoLoRe} generates a $4096^3$ cell box as a compromise between resolution and memory usage, given the large volume that we must cover in order to realistically represent a \Lya\ forest survey. Generating approximately $7.5\mathrm{M}$ QSOs (across the whole sky) and drawing subsequent skewers produces a dataset sufficient for a DESI-like survey, allowing for a significant degree of flexibility in the final survey strategy and number densities. The computational cost of producing one such dataset is relatively low, provided suitable multi-node, multi-core computational facilities are available. Running \texttt{CoLoRe} using the input data and options specified in \secref{subsec:Generating realisations} in parallel across 32 Haswell compute nodes (each with 32 cores and 128GB of memory) on the National Energy Research Scientific Computing Centre's \textit{Cori} machine requires approximately $18$ minutes to run, equivalent to approximately 300 CPU hours. The large number of nodes is necessary to improve the speed of the code and to satisfy its memory requirements --- a total of approximately 920 GB is needed for each run of this size. If such facilities are not available, then the box size must be reduced or the resolution lowered.

The precise requirements for running \texttt{LyaCoLoRe} depend strongly on the exact choices of input options. As an example, converting $800\mathrm{k}$ skewers --- similar to the number that will be observed by DESI --- from \texttt{CoLoRe}'s Gaussian output to realistic transmission skewers including RSDs (though not thermal broadening effects) requires only $4$ minutes when spread across the same 32 nodes mentioned previously. If such computational facilities are not available, then running \texttt{LyaCoLoRe} is still possible as its memory requirements are much lower than \texttt{CoLoRe}.

A very small test dataset of 1000 skewers is available within the \texttt{LyaCoLoRe} repository. It is straightforward to run \texttt{LyaCoLoRe} on this data on any standard laptop to generate sample skewers or to explore the functionality of the code.

\section{Parameter tuning}
\label{sec:Parameter tuning}

A number of parameters are defined in the various transformations described in \secref{subsec:LyaCoLoRe}, namely $n$, $k_1$, $\sigma_\epsilon(z)$, $\tau_0(z)$ and $\alpha(z)$ (see equations \eqref{extra power}, \eqref{gaussian} and \eqref{FGPA} for definitions). These are all free parameters, and we would like to be able to choose their values so that our final skewers have particular properties. Specifically, we aim to match the 1D power spectrum $P_\mathrm{1D}(k,z)$, mean transmitted flux fraction $\bar{F}(z)$ and large-scale bias $b_{\delta,F}(z)$ (as defined in equation \eqref{eq:Power spectrum of flux}) to literature values. Ignoring RSDs and the shape of the 1D power spectrum would allow the problem to be treated analytically, but unfortunately such simplifications are unrealistic. As such, it is not obvious how to choose our parameters correctly, and a more complex process is necessary.

We aim to solve this problem via a minimisation process. We first define a function that we will aim to minimise, and which takes the following steps:
\begin{enumerate}
    \item Generate sample skewers in $F$ corresponding to a given set of parameter values using the methods described in \secref{subsec:LyaCoLoRe}.
    \item \label{Tuning step: measure} Measure the 1D power spectrum, mean flux and large-scale bias of these skewers at a selection of redshift values.
    \item \label{Tuning step: quantify errors} Evaluate the deviation of each measurement at each redshift from literature results.
    \item \label{Tuning step: sum errors} Quantify this deviation with a single number.
\end{enumerate}
In step \ref{Tuning step: measure}, we measure $P_\mathrm{1D}$ and $\bar{F}$ straightforwardly, excluding cells that sit at a rest frame wavelength above $1200\ \angstrom$. We measure $b_{\delta,F}$ by calculating the response of $\bar{F}$ to a small deviation in the average density field: $b_{\delta,F} = (1/\bar{F})\ \mathrm{d}\bar{F}/\mathrm{d}\delta$~\cite{McDonald:2003ApJ...585...34M}. The literature values referred to in step \ref{Tuning step: quantify errors} are the fitting function from the BOSS DR9 $P_\mathrm{1D}$ measurement from~\cite{Palanque-Delabrouille:2013A&A...559A..85P}, the fitting function of the mean flux measurement from~\cite{Becker:2013MNRAS.430.2067B} and the bias value and redshift evolution determined by the BOSS DR12 combined \Lya\ auto- and cross-correlation analysis in~\cite{duMasdesBourboux:2017A&A...608A.130D}. Using these literature results as targets, we compute a weighted error for each measurement at each redshift value. When computing the error on the $P_\mathrm{1D}$, we prioritise the low-$k$ modes by using a $k$-dependent error weighting. For $k<0.02\ \mathrm{s\ km}^{-1}$, this is proportional to $1/(1+(k/k_0)^2)$ where $k_0=0.01\ \mathrm{s\ km}^{-1}$. This ensures the modes most relevant for a BAO analysis --- those with $k\lessapprox0.005\ \mathrm{s\ km}^{-1}$ \cite{McDonald:2007PhRvD..76f3009M,McQuinn:2011MNRAS.415.2257M} --- are prioritised over less important, high-$k$ modes. Beyond $k=0.02\ \mathrm{s\ km}^{-1}$, we ignore any errors as our finite cell size makes it unreasonable to expect realistic power at these scales, and these modes were not measured by BOSS. We sum the errors in quadrature over all $k$-modes using this weighting to produce an overall error on $P_\mathrm{1D}$. In step \ref{Tuning step: sum errors}, the errors on each measurement at each redshift value are summed in quadrature, and a single number produced. This number quantifies how well a given parameter set is able to produce realistic data, as measured by our specified properties. A standard minimisation routine can then be used to minimise it over the space of input parameters. We use \texttt{Minuit}~\cite{James:1975CoPhC..10..343J}, as implemented by the python module \texttt{iminuit}\footnote{Publicly available at \href{https://github.com/iminuit/iminuit}{https://github.com/iminuit/iminuit}.} to do so.

We introduce a number of simplifications to improve the speed of the minimisation. We assume that $\log \tau_0$ and $\log \sigma_\epsilon$ follow the functional form:
\begin{equation}
    \log(X) = A_0 + A_1 \log [(1+z)/(1+z_0)],
\end{equation}
where $z_0 = 3.0$ and the $A_i$ are scalar parameters. In the case of $X=\tau_0$, we fix $A_1=4.5$ \cite{Seljak:2012JCAP...03..004S}. Further, we assume that $\alpha(z)$ takes a constant value of 1.65 across redshifts \cite{McDonald:2001ApJ...562...52M} (equivalent to a value of $\gamma = (2-\alpha)/0.7 + 1$ of 1.5, in reasonable agreement with the literature~\cite[e.g.][]{Ricotti:2000ApJ...534...41R,Hiss:2018ApJ...865...42H}). With these simplifications, we end up with a 5-parameter minimisation problem: 1 parameter describing the normalisation of $\tau_0(z)$; 2 describing the normalisation and $z$-dependence of $\sigma_\epsilon (z)$; and 2 describing the shape of the 1D power of the small scale fluctuations ($n$ and $k_1$). At each call of the routine, we produce sample skewers at a point in parameter space, and compute their 1D power spectra, mean flux and bias parameter values in 7 redshift bins of width $\Delta z=0.2$ centred at points evenly spaced between $z=2.0$ and 3.2. We run this procedure using $\sim55,000$ skewers to obtain an initial estimate, and increase this to $\sim220,000$ skewers in order to fine tune the optimisation.

We also introduce a parameter $a_v$ by which we multiply the velocities in our skewers in order to match the amount of anisotropy in the clustering of the \Lya\ forest to literature values. This is not because the velocities from \texttt{CoLoRe} are incorrect --- when using \texttt{CoLoRe}'s unmodified velocities, we obtain the correct level of anisotropy in the QSO auto-correlation (see Appendix~\ref{sec:Quasar auto-correlation}) --- but is a result of the approximations in our recipe to estimate $F$. These approximations define the relationship between \texttt{CoLoRe}'s initial Gaussian field and our final flux skewers, and thus their inherent assumptions will affect the large-scale flux biases. While the density bias $b_\delta$ is matched to literature values in our tuning process, this is not the case for the RSD parameter $\beta$ (defined in \secref{subsec:Fitting the correlation functions}). As such, the somewhat crude nature of our approximations results in an unrealistic value of $\beta$, and we must introduce our parameter $a_v$ in order to correct for this shortcoming. Improving on our approximations by using more detailed modelling \citep[e.g.][]{Irsic:2018JCAP...04..026I} might allow us to avoid introducing $a_v$, but that is beyond the scope of this work. We fix $a_v=1.3$ when tuning; it is computationally costly to leave it free as a change in $a_v$ requires re-computation of the RSD weights matrix $W_{ij}$ (see \secref{subsubsec:Adding redshift-space distortions}). The value is chosen on an ad hoc basis to match approximately the RSD parameter $\beta$ measured from BOSS DR12 data~\cite{duMasdesBourboux:2017A&A...608A.130D}.

\begin{figure}
\centering
\includegraphics[height=0.30\textheight]{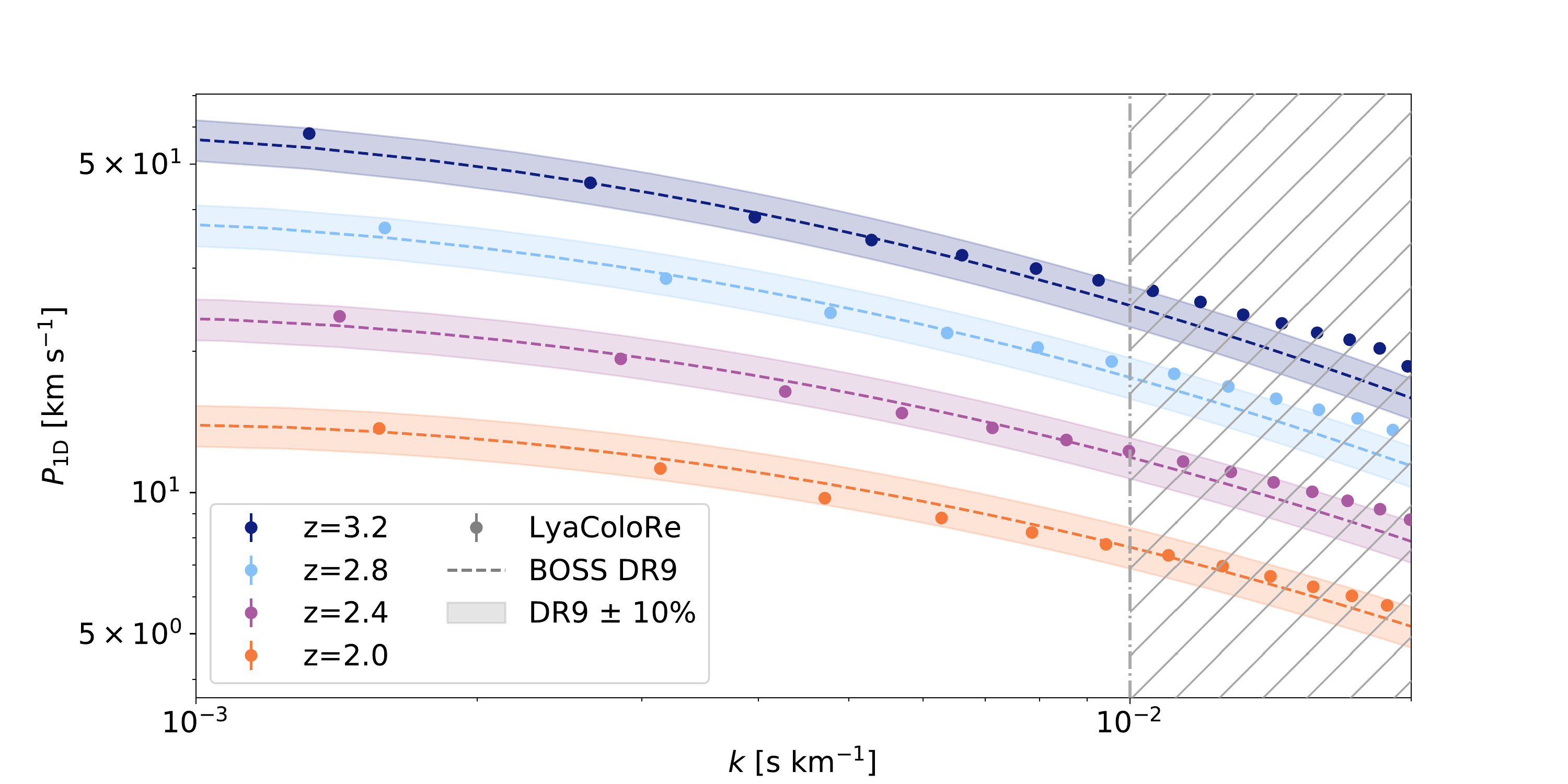}
\caption{The 1D power spectrum as measured from one realisation of \texttt{LyaCoLoRe} mocks. The tuning process aims to match the measured $P_\mathrm{1D}$ to that from BOSS DR9 data~\cite{Palanque-Delabrouille:2013A&A...559A..85P} for $k$-modes that affect BAO analysis, as described in detail in \secref{sec:Parameter tuning}. Modes to the left of the dot-dash line at $k=0.01~\mathrm{s\ km}^{-1}$ are the most important in this respect \cite{McDonald:2007PhRvD..76f3009M,McQuinn:2011MNRAS.415.2257M}, and these all lie within 10\% of our target $P_\mathrm{1D}$, as indicated by the shaded areas. Modes to the right of the dot-dash line are not important in the context of BAO, and are not prioritised in our tuning procedure.}
\label{fig:Pk1D_k0_vs_k_flux}
\end{figure}

The final values of the transformation parameters are $\log[\tau_0(z)]= 1.48 + 4.5 \log x$, $\alpha(z) = 1.65$, $\log[\sigma_\epsilon(z)] = 6.02 + 0.276 \log x$, $n = 0.732$, $k_1=0.0341$ and $a_v=1.3$, where $x=[(1+z)/(1+z_0)]$ and numerical values are rounded to three significant figures where appropriate. These are the default values used by \texttt{LyaCoLoRe}. The tuning process is effective, matching literature values of $P_{\mathrm{1D}}$, $\bar{F}$ and $b_{\delta,F}$ to within 10\% at almost all relevant $k$-modes and $z$ values. As an example, the $P_{\mathrm{1D}}$ measured across $\sim 7.5$M skewers is shown in Figure \ref{fig:Pk1D_k0_vs_k_flux}. We only plot 4 redshift bins and a limited number of $k$-modes here for clearer visualisation.

\section{Verifying the mocks}
\label{sec:Verifying the mocks}

The primary motivation for creating the \texttt{LyaCoLoRe} mocks is to provide realistic sets of test skewers for BAO analyses from \Lya\ forest surveys. Evidently then, it is important to verify that the fundamental physical quantities studied by such analyses are correctly reproduced in the mock datasets. We thus seek to test that the BAO signal is present and unbiased in our mock datasets. \secref{subsec:Generating realisations} describes the inputs we use in generating a collection of mock datasets; \secref{subsec:Measuring correlation functions} explains how we measure the correlation functions from each realisation, taking the skewers in $F$ directly from \texttt{LyaCoLoRe}'s output; and finally \secref{subsec:Fitting the correlation functions} shows how we fit to a model. We do not visually compare the correlation functions measured from mocks to those from data, since our mock measurements are not affected by distortions from continuum fitting. Instead we compare fitted parameter values in order to assess the performance of our mock datasets.

\subsection{Generating realisations}
\label{subsec:Generating realisations}

The input power spectrum that we use in step 1 of \secref{subsec:CoLoRe} is generated by the Boltzmann solver \texttt{CAMB}~\cite{Lewis:2000ApJ...538..473L} using the \textit{Planck} Collaboration's 2015 parameters for a flat, $\Lambda$CDM cosmology~\cite[see column 1 of Table 3 in][]{PlanckCollaboration:2016A&A...594A..13P}. We generate the field in a box of $4096^3$ cells, stipulating that this covers a redshift range $0.0\leqslant z \leqslant3.79$: a volume large enough to contain a DESI-like survey. This results in a grid of total size $\sim(9.8\ \mathrm{Gpc/}h)^3$, with each cell $\sim(2.4\ \mathrm{Mpc/}h)^3$ in dimensions. The QSO number density function is based on estimates from SDSS-III data in Stripe 82~\cite{Palanque-Delabrouille:2016A&A...587A..41P}. This is considered to represent an optimistic estimate of the photometric capability of targeting for DESI, and results in $\sim 3.7$M QSOs\footnote{This is lower than the 7.5M quoted in \secref{subsec:Computational requirements} as we no longer require the previously mentioned flexibility to adapt to different observing strategies in our realisations, and thus can reduce the QSO number density to more realistic values (approximately 59 QSOs per square degree).} above $z=1.8$ across the whole sky. We use as an input QSO bias the fitting function defined in equation 19 of \cite{GontchoAGontcho:2018MNRAS.480..610G}, which is based on clustering measurements from the BOSS DR12 QSO sample \cite{Laurent:2016JCAP...11..060L}. When running \texttt{LyaCoLoRe}, we use a cell size of 0.25 Mpc/$h$, and tune the parameters of our transformations according to the methods described in \secref{sec:Parameter tuning}.

For the purposes of this work we generate 10 such realisations, each with unique random seeds, and stack our results in order to test \texttt{LyaCoLoRe} as stringently as possible. This is approximately equivalent to 30 times the final number of \Lya\ QSOs with $z\geq2.1$ that will be observed by DESI. It is worth noting that the signal to noise ratio will be significantly greater than 30 times that of DESI, as our skewers of $F(\lambda)$ do not include any instrumental noise, nor do they require any continuum fitting (as mentioned in \secref{sec:Making the mocks}).

\subsection{Measuring correlation functions}
\label{subsec:Measuring correlation functions}

We test the BAO signal in our mock realisations in the standard way, by measuring correlation functions using the contrast in flux transmission:
\begin{equation}
    \delta_F(\lambda) = \frac{F(\lambda)}{\bar{F}(\lambda)} -1,
\end{equation}
where $\bar{F}(\lambda)$ is the mean value of $F(\lambda)$ in each pixel over all skewers for which that cell corresponds to rest-frame wavelength $\lambda_r \in [1040,1200]\ \angstrom$. The skewers of $F(\lambda)$ are taken straight from the processes described in \secref{sec:Making the mocks}, with no further steps such as addition of continua or instrumental noise. This allows us to test the methods of \secref{sec:Making the mocks} to as high a degree of precision as possible, but consequently our covariance matrices may not necessarily be representative of true measurements.

We would like to measure the 3D \Lya\ auto-correlation and the 3D \Lya-QSO cross-correlation, the standard measurements made by recent \Lya\ BAO analyses from BOSS and eBOSS. Both are estimated using the \texttt{Package for IGM Cosmological-Correlations Analyses} (\texttt{picca})\footnote{Publicly available at \href{https://github.com/igmhub/picca}{https://github.com/igmhub/picca}.}. We measure these correlations separately in 3,072 \texttt{HEALPix}~\cite{Górski:2005ApJ...622..759G} pixels on the sky for each of the 10 realisations, and treat the resultant measurements as a set of 30,720 independent subsamples. In order to compute the correlation functions more quickly, we rebin pixels in our final transmission skewers into larger pixels of width $3\times10^{-4}$~log(\AA) in log-wavelength. This enables us to use a larger number of skewers and thus reduce our errors, without compromising the large-scale properties of the correlations or incurring large computational costs.

Our computation of the 3D \Lya\ auto-correlation follows that of recent \Lya\ forest BAO analyses \cite{Bautista:2017A&A...603A..12B,deSainteAgathe:2019A&A...629A..85D}. We first define a grid of bins in parallel and perpendicular separation between pairs of pixels --- $r_\parallel$ and $r_\perp$ respectively --- where each bin is 4~Mpc/$h$ $\times$ 4~Mpc/$h$ in size, and the maximum separation is 200~Mpc/$h$ in each direction. Pixel pairs are assigned to one of these bins by using a fiducial cosmology to convert from wavelength and angular separations to comoving distances parallel and perpendicular to the line-of-sight. The correlation is then computed as a weighted sum of products of pixel pairs of $\delta_F$ within each bin. We restrict ourselves to include only contributions from the \Lya\ absorption in the \Lya\ region, ignoring delta pixels outside the rest-frame wavelength range [1040,1200]~\AA. The covariance matrix is estimated straightforwardly by calculating the scatter between our set of 30,720 subsamples.

The 3D \Lya-QSO cross-correlation is also computed in line with recent analyses of BOSS and eBOSS data~\cite{duMasdesBourboux:2017A&A...608A.130D,Blomqvist:2019A&A...629A..86B}, as a weighted sum of pixels of $\delta_F$ within bins of parallel and perpendicular separation. We use the same bin size as in the auto-correlation, but are able to extend our minimum value of $r_\parallel$ to $-200$~Mpc/$h$ as the pixel-pixel pair symmetry of the auto-correlation is not present in the pixel-QSO pairs of the cross-correlation. As for the \Lya\ auto-correlation, we restrict the rest-frame wavelength range of our $\delta_F$ pixels to [1040,1200]~\AA, and we estimate our covariance matrix from the scatter between our 30,720 subsamples.

\subsection{Fitting the correlation functions}
\label{subsec:Fitting the correlation functions}

Having measured the 3D \Lya\ auto- and \Lya-QSO cross- correlations, we fit a model to our measurements to obtain the location of the BAO peak and check that no significant shift has been introduced. We also seek to measure the bias parameters of our tracers: \Lya\ flux $F$ and QSOs. These are defined by the relationship between the power spectra of the tracers, $P_F(\mathbf{k})$ and $P_\mathrm{QSO}(\mathbf{k})$, and the power spectrum of dark matter $P(\mathbf{k})$ \cite{Kaiser:1987MNRAS.227....1K}:
\begin{equation}
\label{eq:Power spectrum of flux}
    P_F(\mathbf{k}) = [b_{\delta,F} + b_{\eta,F} f \mu^2]^2 P(\mathbf{k}),
\end{equation}
\begin{equation}
\label{eq:Power spectrum of QSOs}
    P_\mathrm{QSO}(\mathbf{k}) = [b_{\delta,\mathrm{QSO}} + f \mu^2]^2 P(\mathbf{k}).
\end{equation}
Here, the large-scale biases of flux and QSOs are $b_{\delta,F}$ and $b_{\delta,\mathrm{QSO}}$. The parameter $b_{\eta,F}$ is the velocity gradient bias of flux, which serves to quantify the effect of RSDs. This is often expressed alternatively using $\beta = f b_{\eta,F} / b_{\delta,F}$. The value of $b_{\eta,QSO}$ is 1 by default as QSOs are conserved under RSDs (unlike $F$) and so it is held fixed \cite{Kaiser:1987MNRAS.227....1K}. The \Lya-QSO cross- power spectrum follows naturally from \eqref{eq:Power spectrum of flux} and \eqref{eq:Power spectrum of QSOs} as:
\begin{equation}
\label{eq:Power spectrum of Lya QSO cross}
    P_{F\times \mathrm{QSO}}(\mathbf{k}) = [b_{\delta,F} + b_{\eta,F} f \mu^2] [b_{\delta,\mathrm{QSO}} + f \mu^2] P(\mathbf{k}).
\end{equation}

We fit a model of the correlation functions to each of the measurements individually, and then to both correlations jointly. We use the same models as recent eBOSS analyses \cite{deSainteAgathe:2019A&A...629A..85D,Blomqvist:2019A&A...629A..86B} but ignore terms relating to systematics not present in our realisations, such as metal absorbers and high column density systems (HCDs). As we do not add continua to our skewers, we need not worry about the distortion of the correlations by the removal of long wavelength modes in the continuum fitting process, as occurs in real analyses. Thus, we do not need to consider distortion matrices, the standard method for taking these effects into account \cite[introduced for the auto- and cross- correlations respectively in][]{Bautista:2017A&A...603A..12B,duMasdesBourboux:2017A&A...608A.130D}. The relevant terms are described using Kaiser models~\cite{Kaiser:1987MNRAS.227....1K}, as described in section 4.1 of \cite{deSainteAgathe:2019A&A...629A..85D} for the \Lya\ auto-correlation, and section 5.1 of \cite{Blomqvist:2019A&A...629A..86B} for the \Lya-QSO cross-correlation. We use the same cosmology as used to generate the input power spectrum of \texttt{CoLoRe} to produce the smooth and peak components of the fiducial model power spectrum.

The fit is carried out leaving free the parameters describing the position of the BAO peak in the perpendicular and parallel directions:
\begin{equation}
    \alpha_\parallel = \frac{D_H(z)/r_d}{[D_H(z)/r_d]_{\rm{fid}}},\ \ \alpha_\perp = \frac{D_A(z)/r_d}{[D_A(z)/r_d]_{\rm{fid}}},
\end{equation}
where $D_H(z)=c/H(z)$, as well as parameters describing the bias and RSDs of the \Lya-forest, $b_{\eta,F}$ and $\beta_F = f b_{\eta,F} / b_{\delta,F}$. We also leave free 2 parameters that describe the smoothing of the model power spectrum in the parallel and perpendicular directions, which help to account for the effects of the low-resolution of our \texttt{CoLoRe} grid. When fitting the \Lya-QSO cross-correlation individually, we fix the value of the QSO bias $b_{\delta,\mathrm{QSO}}$ to the input value in order to avoid degeneracies, though when we fit jointly with the \Lya\ auto-correlation we are able to leave it free.

Having defined our models, the fits are then carried out using \texttt{picca}. We fit only on separations $40<r$~[Mpc/$h]<160$ as the lognormal density approximation used in both \texttt{CoLoRe} and \texttt{LyaCoLoRe} begins to break down on scales smaller than this, and we are not able to fit the shape of the correlation function well at these separations. Further, the QSOs cannot be expected to be correctly clustered on the smallest scales due to the low-resolution of the \texttt{CoLoRe} box. To determine an effective redshift of our measurements, we consider pixel-pixel/pixel-QSO pairs which fall in bins $A$ which satisfy $80<r_A\ [\mathrm{Mpc}/h]\ < 120$, i.e. the bins that cover the BAO peak. The value of $z_\mathrm{eff}$ is then given by a weighted average of the redshifts of pairs in these bins.

\begin{figure}
\centering
\includegraphics[height=0.34\textheight]{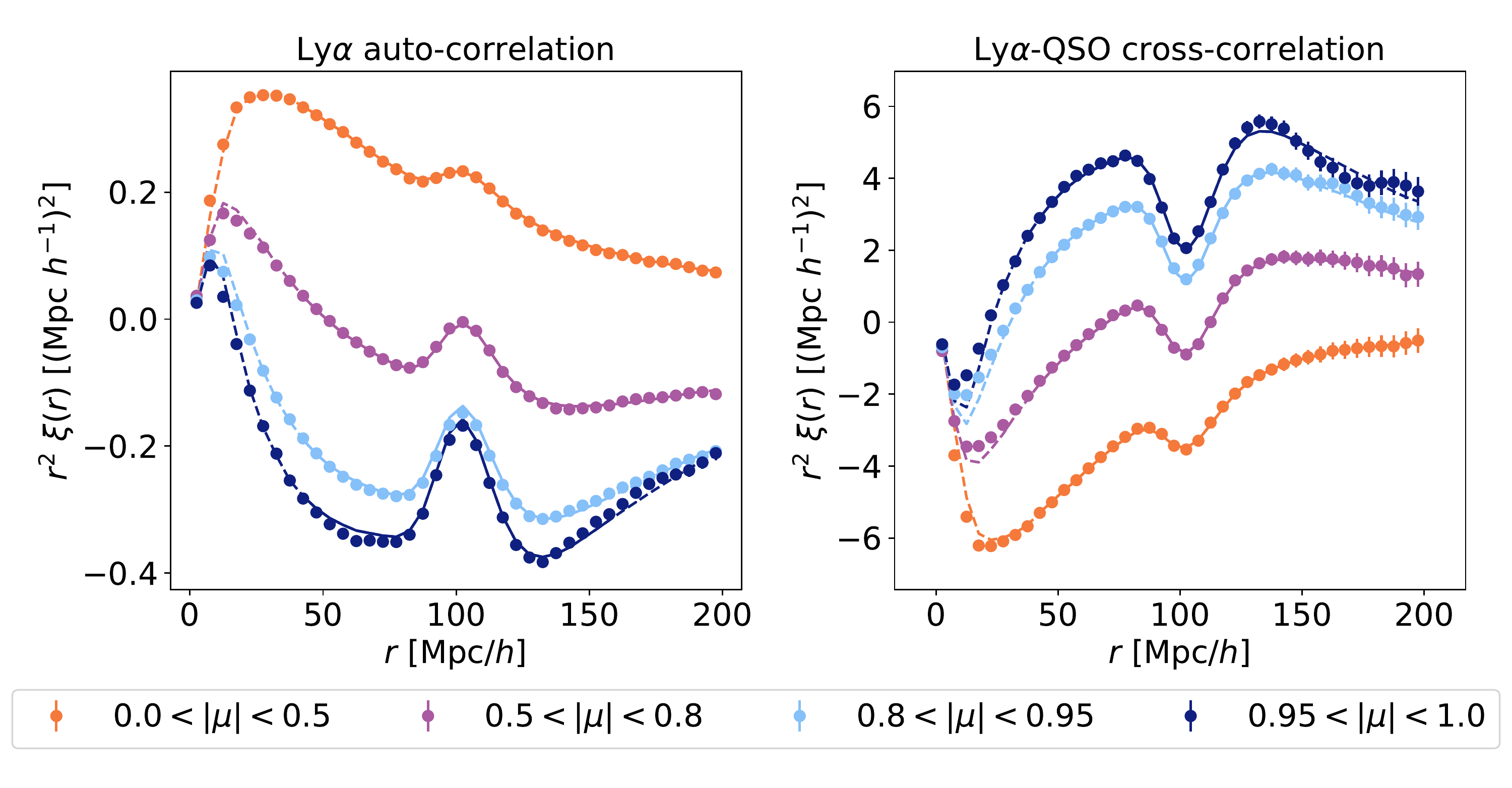}
\caption{The correlation functions measured from 10 realisations of \texttt{LyaCoLoRe} datasets combined, and the best fit lines with parameters as described in the third column of Table \ref{table:fit parameters}. The left panel shows the \Lya\ auto-correlation, while the right panel shows the \Lya-QSO cross-correlation. Each plot panel shows the same 4 bins in $|\mu|=|r_\parallel|/r$. Note that the correlations presented here do not have any distortion from continuum-fitting and so should not be visually compared with the equivalent plots from recent BOSS/eBOSS data.}
\label{fig:corr_plot}
\end{figure}

The measured \Lya\ auto- and \Lya-QSO cross-correlations are shown in the left and right panels of Figure \ref{fig:corr_plot} respectively, along with the model from the combined fit\footnote{Note that error bars are present for all points, but are often exceedingly small and thus obscured by the points themselves.}. We plot the correlations as $\xi(r)$ in bins of $|\mu| = |r_\parallel|/r$, where $|\mu|$ close to 0 indicates correlations close to perpendicular to the line of sight, and $|\mu|$ close to 1 indicates correlations close to parallel to the line of sight. The model appears to be a good fit to the measurement on the scales that we fit over, and the BAO peak is correctly placed. The two measurements deviate slightly from the model either side of the BAO bump in the highest $|\mu|$ bin, but this deviation is very small and is noticeable due to the extremely small error bars on our measurements.

The parameters from the individual and combined fits are shown in Table \ref{table:fit parameters}. The $\alpha$ parameters for each fit are all consistent with 1 to within $1\sigma$. Any deviation in the $\alpha$ parameters from 1 is certainly less than 0.2\%, and so can be considered insignificant in the context of a DESI-like survey. Thus, the mock production pipeline up to this stage can be said to introduce no clear systematic bias within the capabilities of a current or near-future instrument.

In order to compare the values of biases and $\beta$s to BOSS DR12 values \cite[table 4 of][]{duMasdesBourboux:2017A&A...608A.130D}, we first use the published functional forms of each parameter's redshift evolution to match the effective redshift of the BOSS DR12 measurements to that of our measurements. Having done so, we find that the two sets of values are very similar, with our measurements all lying within the 1$\sigma$ errors on the BOSS DR12 values. In particular, the values of $b_{\delta,F}$ in each of our fits are almost identical to the BOSS DR12 value, demonstrating the effectiveness of our tuning of this parameter (see \secref{sec:Parameter tuning}). We do not compare the value of $\beta_{\mathrm{QSO}}$ to BOSS DR12 measurements as our input QSO bias takes a different value at this redshift. However, the value of $b_{\delta,\mathrm{QSO}}$ deduced from the joint fit is consistent with the input value (as shown at the bottom of the column showing the \Lya-QSO only fit). As such, we can consider the mocks to fulfil the basic criteria required of them, and thus they appear sufficient for a DESI-like survey. We do not assess the $\chi^2$ of the fits as we do not expect our covariance matrices to be representative of those one would expect from a real survey given the lack of noise in our skewers.

\begin{table}
\centering
\begin{tabular}{l|ccc|c}
\hline
\hline
                        & \multicolumn{3}{|c|}{\phm\texttt{LyaCoLoRe}\phc}                                  & \phm BOSS DR12                \\
\hline
Parameter               & \phm\Lya\phc              & \phm\Lya-QSO\phc          & \phm\Lya\ $+$\phc         & \phm\Lya\ $+$                 \\
name                    &                           &                           & \phm\Lya-QSO\phc          & \phm\Lya-QSO                  \\
\hline
\hline
$\alpha_\parallel$      & \phm$1.000 \pm 0.002$\phc & \phm$1.001 \pm 0.002$\phc & \phm$1.000 \pm 0.001$\phc &                               \\
$\alpha_\perp$          & \phm$0.998 \pm 0.002$\phc & \phm$1.000 \pm 0.002$\phc & \phm$0.999 \pm 0.001$\phc &                               \\
$b_{\eta,F}$            & $-0.204\pm 0.0004$        & $-0.201 \pm 0.0009$       & $-0.203 \pm 0.0004$       & $-0.206 \pm 0.012$            \\
$\beta_F$               & \phm$1.627 \pm 0.008$\phc & \phm$1.624 \pm 0.012$\phc & \phm$1.624 \pm 0.007$\phc & \phm$1.650 \pm 0.081$         \\
$\beta_\mathrm{QSO}$    &                           & $0.261$                   & \phm$0.261 \pm 0.0007$    &                               \\
\hline
$b_{\delta,F}$          & $-0.121 \pm 0.0006$       & $-0.120 \pm 0.0009$       & $-0.121 \pm 0.0005$       & $-0.121 \pm 0.004$            \\
$b_\mathrm{QSO}$        &                           & $3.701$                   & \phm$3.700 \pm 0.009$\phc &                               \\
\hline
\hline
\end{tabular}
\caption{Parameters from model fits of the \Lya\ auto-correlation and \Lya-QSO cross-correlation functions measured from 10 realisations of \texttt{LyaCoLoRe} mocks combined. The relevant results from the BOSS DR12 combined fit~\cite{duMasdesBourboux:2017A&A...608A.130D} --- those to which our values of $b_{\delta,F}$ and $b_{\eta,F}$ are tuned --- are presented in the rightmost column at the same effective redshift as our measurements. The parameters in the first segment of the table are those used in the minimisation process which determines the best fit to our correlations, while those in the second segment are calculated subsequently.}
\label{table:fit parameters}
\end{table}

\section{Adding secondary astrophysical effects}
\label{sec:Adding secondary astrophysical effects}

A key purpose of creating mock datasets is to quantify the impact of secondary astrophysical effects on our measurements so that we may assess any biases that they could induce in our cosmological inference. When generating realisations of the synthetic data, we may choose to add or not to add different effects to each realisation, or to vary the strength of a given effect across a range of values. The resultant impact on BAO measurements can then be quantified. In \Lya\ forest analyses, two of the most pertinent effects are the presence of high column density systems (HCDs) and additional absorption transitions. \texttt{LyaCoLoRe} is able to compute both of these effects, and the methods it uses to do so are described in \secref{subsec:Adding HCDs} and \secref{subsec:Including additional absorption transitions} respectively (alternative implementations of these effects are also possible). Once computed, \texttt{LyaCoLoRe} stores skewers of metal absorption and a table of HCDs in its output. These can then be added to the \Lya\ skewers during subsequent stages of the pipeline by packages such as \texttt{desisim}.

We do not present here a full study of the effects of HCDs and additional transitions on a BAO analysis. Rather, we simply illustrate in \secref{subsec:Testing astrophysical effects} that their implementations within \texttt{LyaCoLoRe} are broadly correct and achieve the correct levels of clustering. We leave the study of these effects as systematics in a BAO analysis for future work.

\subsection{Adding HCDs}
\label{subsec:Adding HCDs}

HCDs occur in particularly dense regions of gas, and contain a number of subclasses determined by HI column density. Typically, we define regions with column density $N_{\mathrm{HI}} > 2\times10^{20}\ \mathrm{cm}^{-2}$ as Damped \Lya\ absorbers (DLAs), and regions with column density $10^{17.2} < N_{\mathrm{HI}} < 2\times10^{20}\ \mathrm{cm}^{-2}$ as Lyman Limit Systems (LLSs) \cite{Wolfe:1986ApJS...61..249W}. In detailed \Lya\ forest analyses, it is important to be able to identify HCDs as their high densities broaden their absorption profiles, impacting on inferred values of $F$ over a significant wavelength range \cite{Font-Ribera:2012JCAP...07..028F,Rogers:2018MNRAS.476.3716R}. Further, HCDs are of scientific interest in and of themselves \cite[e.g.][]{Pettini:1997ApJ...486..665P,Prochaska:2002PASP..114..933P,Padmanabhan:2016MNRAS.458..781P,Pérez-Ràfols:2018MNRAS.480.4702P,Pérez-Ràfols:2018MNRAS.473.3019P}. As such, being able to add HCDs to our mocks is important in maximising their realism.

We first determine potential HCD locations by computing a threshold value of the Gaussian field, set by an input bias $b_\mathrm{HCD}(z)$. In our realisations, we choose $b_\mathrm{HCD}(z)=2.0$ to be constant with redshift, and in line with \cite{Pérez-Ràfols:2018MNRAS.473.3019P}. This picks out peaks in the field that are sufficiently dense to host an HCD. We then Poisson sample the potential locations according to an input number density $n_\mathrm{HCD}(z)$. This number density is imported from the default model of the IGM physics package \texttt{pyigm}\footnote{Publicly available at \href{https://github.com/pyigm/pyigm}{https://github.com/pyigm/pyigm}.} \cite{j_xavier_prochaska_2017_1045480}, which is fitted to a selection of literature results \cite[summarised in Table 1 of][]{Prochaska:2014MNRAS.438..476P}. The sampling is carried out before adding small-scale power (\secref{subsubsec:Adding small-scale power}), as we would like the HCDs to correlate with the 3D fluctuations rather than the 1D extra power. A column density is then allocated to each HCD using a given redshift distribution --- again from the default model of \texttt{pyigm} --- and a radial velocity is determined using \texttt{CoLoRe}'s output. The resulting catalogue of HCDs can then be interpreted by a package such as \texttt{desisim}, which is able to calculate the absorption profile of the HCD using a Voigt template, and insert it into the final spectrum.

\subsection{Including additional absorption transitions}
\label{subsec:Including additional absorption transitions}

As with HCDs, absorption from additional transitions are an important level of detail to add to our mocks and are of significant scientific interest~\citep[e.g.][]{Pieri:2014MNRAS.441.1718P,Blomqvist:2018JCAP...05..029B,GontchoAGontcho:2018MNRAS.480..610G,duMasdesBourboux:2019ApJ...878...47D}. Additional transitions have a rest-frame absorption wavelength different to that of \Lya, and so absorption from gas at the same redshift appears at different observed wavelengths in spectra. Conversely, absorption from two different transitions can appear at the same observed wavelength even though the regions of gas hosting the absorbers are far apart physically. As a result, the presence of such absorption transitions acts to contaminate our measurements of \Lya\ flux, and thus our correlation functions and resultant BAO measurements. Such transitions include Lyman-$\beta$ (\Lyb), as well as from silicon, oxygen and carbon gas, for example.

Similar to the method to add HCDs described in \secref{subsec:Adding HCDs}, it would also be reasonable to place additional absorption transitions using a Poisson-sampled ``density-peak'' approach, as metals are typically produced in high density regions of the universe. However, we choose to follow the methods of previous works~\cite{Slosar:2011JCAP...09..001S,Bautista:2015JCAP...05..060B}, assuming that the optical depth of each additional transition is proportional to that of the \Lya\ absorber. In the context of these mocks, the most important feature of these additional transitions that we seek to replicate is the strength of their 3D clustering, as it is this that will quantify any impact upon BAO measurements. In order to do so, we simply require an absorption strength (relative to \Lya) and a rest frame wavelength for each additional transition that we wish to include. Having calculated the skewers of optical depth in real space, we scale them differently for each absorption transition according to the transition's relative strength. For an additional transition $X$, we obtain $\tau_X = A_X \tau_{\alpha}$, where $A_X$ is the relative strength and $\tau_\alpha$ is the \Lya\ optical depth as defined in equation \eqref{FGPA}. We then apply RSDs (using the same weights matrix $W_{ij}$ as for \Lya), and convert to $F_X(\lambda)$ separately for each $X$ according to its rest frame wavelength. For each line of sight, the separate $F_X(\lambda)$ skewers are then interpolated onto a common wavelength grid and are combined multiplicatively.

This method ensures that RSDs are correctly applied to each additional absorption transition, and we may tune the absorption strength in order to achieve the correct large-scale bias --- and thus the correct 3D clustering --- for each transition. A small selection of additional transitions and their relative strengths are shown in Table \ref{table:metal_absorbers}. These are the transitions most important to a \Lya\ BAO analysis, though further transitions can be added straightforwardly if needed. These strengths have been tuned to approximately match bias values presented in the literature \cite{Bautista:2017A&A...603A..12B,duMasdesBourboux:2017A&A...608A.130D,deSainteAgathe:2019A&A...629A..85D}.

\begin{table}
\centering
\begin{tabular}{l|cc}
\hline
\hline
Name & Rest frame & Relative\\
 & wavelength [\AA] & absorption strength \\
\hline
\hline
\Lya\	        &1215.67    &$1.0$\\
\hline
\Lyb	        &1025.72    &$0.1901$\\
\hline
SiII (1260)     &1260.42    &\phc\phc$3.542\times10^{-4}$\\
SiIII (1207)    &1206.50    &\phc$1.8919\times10^{-3}$\\
SiII (1193)     &1193.29    &\phc$9.0776\times10^{-4}$\\
SiII (1190)     &1190.42    &$1.28478\times10^{-4}$\\
\hline
\hline
\end{tabular}
\caption{Details of a small selection of additional absorption transition that can be used in \texttt{LyaCoLoRe}. The absorption strength for each absorber $X$ has been tuned to match approximately the bias value $b_{\delta,X}$ found in literature \cite{Bautista:2017A&A...603A..12B,duMasdesBourboux:2017A&A...608A.130D,deSainteAgathe:2019A&A...629A..85D}. It is possible to add more absorbers straightforwardly, but the absorption strengths have not been calibrated beyond those listed above. These absorbers are those included in the skewers from which the correlation function in Figure \ref{fig:corr_plot_systematics} is measured.}
\label{table:metal_absorbers}
\end{table}

\subsection{Testing astrophysical effects}
\label{subsec:Testing astrophysical effects}

We assess the methods of \secref{subsec:Adding HCDs} and \secref{subsec:Including additional absorption transitions} by first computing the 3D \Lya-HCD cross-correlation. The methods used to do so are largely the same as used to compute the 3D \Lya-QSO cross-correlation, as described in \secref{subsec:Measuring correlation functions}. One significant difference is that we restrict the HCDs in our calculation of the \Lya-HCD cross-correlation to lie in the rest frame wavelength range $[1040,1100]$~\AA, far from the background QSO. This restriction is necessary to prevent the correlation between \Lya\ flux and QSOs from significantly affecting our measurements close to the line of sight, as is discussed further in Appendix~\ref{sec:The Lya-HCD cross-correlation}. An effect can still be seen in the two $\mu$-bins closest to the line of sight at large values of $r$, though this is mostly beyond the fitted range and so we are still able to measure the degree of clustering in the HCDs well. Future studies may prefer to model this effect in order to avoid reducing the HCD catalogue in this way, but such work is beyond the scope of this analysis. As in section \secref{sec:Verifying the mocks}, we measure correlations on each of our 10 realisations, and combine the measurements.

\begin{figure}
\centering
\includegraphics[height=0.34\textheight]{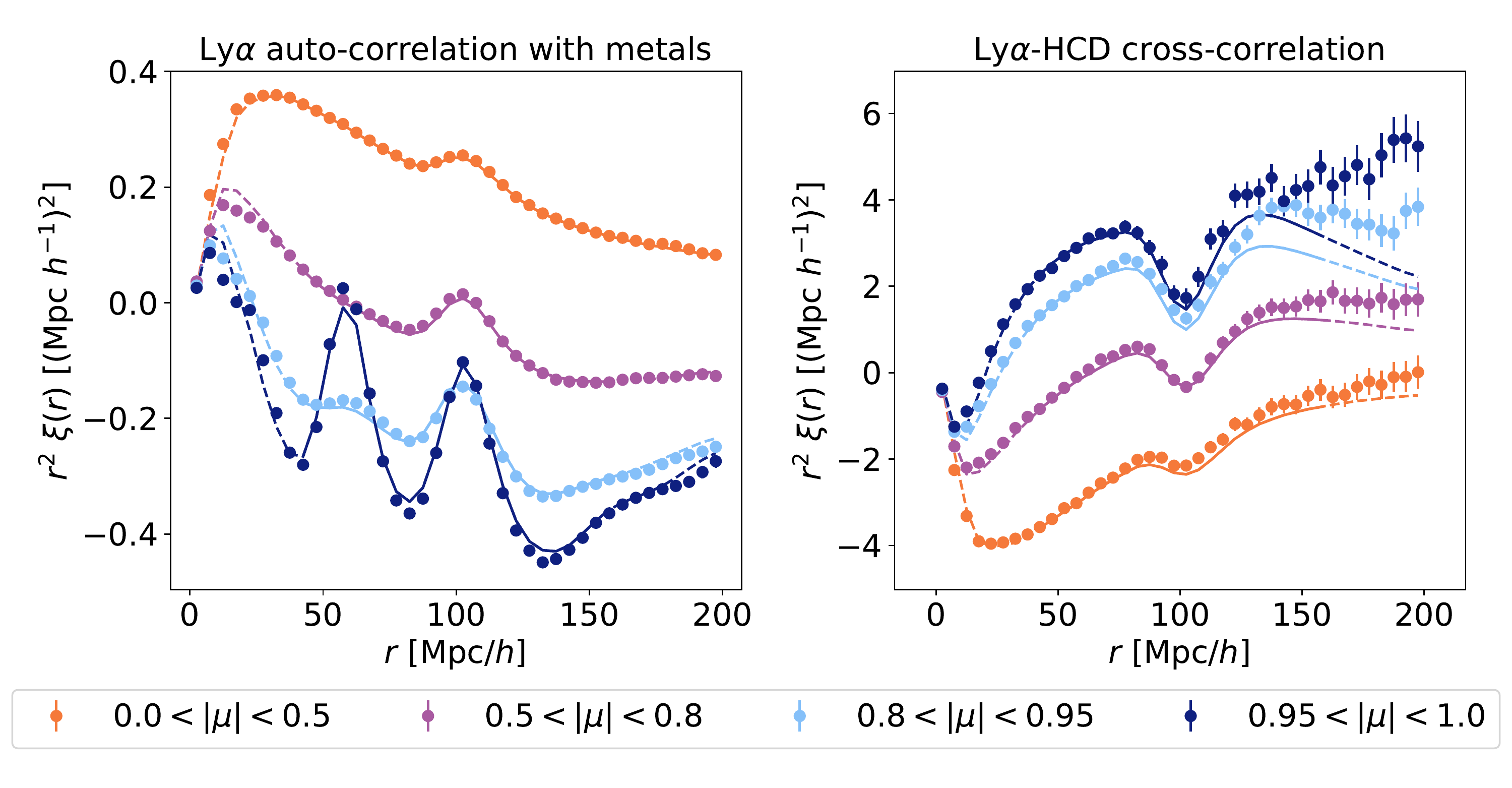}
\caption{The correlation functions measured from 10 realisations of \texttt{LyaCoLoRe} combined, demonstrating the additional astrophysical effects that can be included in its skewers. The left panel shows the flux auto-correlation measured from skewers including metal absorption, from which we measure the metal absorber biases presented in Table \ref{table:metal biases}. The right panel shows the \Lya-HCD cross-correlation. The subtleties of this measurement --- including the discrepancy at large-$r$ --- are discussed in Appendix \ref{sec:The Lya-HCD cross-correlation}. Note that the correlations presented here do not have any distortion from continuum-fitting and so should not be visually compared with the equivalent plots from recent BOSS/eBOSS data.}
\label{fig:corr_plot_systematics}
\end{figure}

The measurement of the \Lya-HCD cross-correlation is shown in the right panel of Figure~\ref{fig:corr_plot_systematics}. Here, we fit for a model in the same way as for the \Lya-QSO cross-correlation. Carrying out a combined fit with the \Lya\ auto-correlation from \secref{sec:Verifying the mocks} allows us to measure the HCD bias $b_{\delta,\mathrm{HCD}}(z_\mathrm{eff}) = 2.26 \pm 0.02$. Strictly, this is not consistent with the redshift-constant input value of $b_{\delta,\mathrm{HCD}} = 2.0$ (as motivated by~\cite{Pérez-Ràfols:2018MNRAS.473.3019P}). There are a number of potential reasons for such a shift, but given the errors on current measurements of $b_{\delta,\mathrm{HCD}}$ from data (approximately 10\% in \cite{Pérez-Ràfols:2018MNRAS.473.3019P}), we do not investigate the agreement further at this stage.

\begin{table}
\centering
\begin{tabular}{l|cccc}
\hline
\hline
                &\multicolumn{3}{c}{bias $\times 10^{3}$}\\
\hline
Absorber        &\texttt{LyaCoLoRe}     &BOSS DR12 &eBOSS DR14\\
\hline
\hline
SiII (1260~\AA)     &\phc$-1.70 \pm 0.04$   &$-1.5 \pm 1.2$     &$-2.5 \pm 1.3$\\
SiIII (1207~\AA)    &$-3.3$      &$-3.3 \pm 1.3$     &$-8.2 \pm 1.0$\\
SiII (1193~\AA)     &\phc$-3.28 \pm 0.03$   &$-3.5 \pm 0.9$     &$-4.6 \pm 1.0$\\
SiII (1190~\AA)     &\phc$-4.55 \pm 0.03$   &$-4.4 \pm 0.9$     &$-5.1 \pm 1.0$\\
\hline
\hline
\end{tabular}
\caption{The biases of the metal absorbers used in our realisations of \texttt{LyaColoRe}, along with the values from BOSS DR12 \cite{Bautista:2017A&A...603A..12B} and eBOSS DR14 \cite{deSainteAgathe:2019A&A...629A..85D} for comparison. The bias of SiIII (1207) is held fixed to the DR12 value as the ``bump'' that it creates in the correlation function is at $r=21$~Mpc/$h$, below the minimum separation of 40~Mpc/$h$ used in our fits. The values from \texttt{LyaColoRe} are all within 1$\sigma$ of those from BOSS DR12, indicating that the absorption strengths used in our realisations (see Table \ref{table:metal_absorbers}) result in the correct levels of large-scale clustering.}
\label{table:metal biases}
\end{table}

We then compute the 3D auto-correlation from skewers of $F$ that include contributions from the additional absorption transitions in Table \ref{table:metal_absorbers} (on top of \Lya\ absorption). The method used to do so is identical to that described for the 3D \Lya\ auto-correlation in \secref{subsec:Measuring correlation functions}. We only include contributions from pixels that lie in the rest-frame wavelength range $[1040,1200]$~\AA, and so we do not include any absorption from the \Lyb\ absorber as its rest-frame wavelength is below this range. As such, from here on we refer to the additional absorption transitions as ``metals''. As in \secref{sec:Verifying the mocks}, we measure correlations on each of our 10 realisations, and combine the measurements.

The measurement of the auto-correlation with metal absorbers is shown in the left panel of Figure \ref{fig:corr_plot_systematics}. By comparison with Figure \ref{fig:corr_plot}, the effect of including these metals in the skewers is clearly significant, particularly in the near-line of sight $0.95<|\mu|<1.0$ bin. Notably, we can clearly see a peak at approximately 55-60~Mpc/$h$ as a result of SiII~(1190~\AA) and SiII~(1193~\AA) absorption, as well as a peak at approximately 21~Mpc/$h$ from SiIII~(1207~\AA). This final peak is not included in our fit as it is below the minimum separation. Less visually obvious, but more important to the BAO analysis, is the effect of absorption from SiII~(1260~\AA), which causes a bump at 105~Mpc/$h$, very close to the BAO peak.

In our fit of this correlation, we model the effect of metal absorbers in the same way as in~\cite{deSainteAgathe:2019A&A...629A..85D}, summing contributions to the model power spectrum from each combination of pairs of absorbers. In Table \ref{table:metal biases}, we show the biases for each of our metal absorbers, as well as the values from BOSS DR12 \cite{Bautista:2017A&A...603A..12B} and eBOSS DR14 \cite{deSainteAgathe:2019A&A...629A..85D} for comparison. The bias of SiIII~(1207~\AA) is held fixed to the DR12 value as the peak that it creates in the correlation function is at $r=21$~Mpc/$h$, below the minimum value used in our fits. The \texttt{LyaColoRe} values sit within 1$\sigma$ of those from BOSS DR12, demonstrating that the levels of clustering given by the absorption strengths in Table \ref{table:metal_absorbers} are similar to those found in data. Of course, each absorption strength can be tuned further so that the bias of the relevant absorber more closely matches any given value.

\section{Summary \& conclusions}
\label{sec:Summary conclusions}
In this work we have presented \texttt{LyaCoLoRe}, a tool for creating mock \Lya\ forest datasets when used in conjunction with a Gaussian random field code such as \texttt{CoLoRe}. We first use \texttt{CoLoRe} to generate skewers from a Gaussian random field, avoiding the use of N-body or hydrodynamical simulations due to the limited volume and high computational cost of such methods. \texttt{LyaCoLoRe} is then able to transform the output into realistic skewers of transmitted flux fraction, with a number of properties defined by an automatic tuning process. The process is computationally efficient, making it suitable for generating large numbers of realisations of mocks with different input data and parameters.

We then demonstrate the effectiveness of \texttt{LyaCoLoRe}'s output, generating a number of skewers equivalent to approximately 30 realisations of DESI and measuring the \Lya\ auto- and \Lya-QSO cross-correlations. Fitting these measurements with an appropriate model gives BAO peak positions that are consistent with the input cosmologies to within 0.2\%, and certainly within the capabilities of an instrument such as DESI. In addition, the biases of the \Lya\ forest and of QSOs are shown to be very similar to those derived from BOSS DR12 data. As such, we conclude that the mock datasets generated by \texttt{LyaCoLoRe} are suitable for the BAO analyses of current and upcoming surveys such as eBOSS and DESI.

Finally, we demonstrate two additional capabilities of the \texttt{LyaCoLoRe} package in adding correlated high column density systems (HCDs) and additional absorption transitions to the skewers. We leave a full analysis on the impact of such features on a BAO analysis to a future work, but demonstrate that the HCDs are clustered approximately correctly on large scales, and that the additional transitions affect the \Lya\ auto-correlation in the expected manner.

Mock datasets such as those generated by \texttt{LyaCoLoRe} are of use to the BAO analyses of \Lya\ forest surveys in a number of ways. They are able to provide robust tests of analysis pipelines, while they can also help in assessing the impact of astrophysical effects --- such as HCDs and additional absorption transitions --- on BAO measurements. Finally, they can be used to provide evidence when making decisions regarding the planning of large surveys, such as in targeting and survey strategy. As such, we hope that \texttt{LyaCoLoRe} will be of use for \Lya\ BAO surveys both present and future.

\section*{Acknowledgements}

We thank Nicolas Busca for valuable discussion, and aiding with the analysis of \texttt{LyaCoLoRe} datasets via \texttt{picca}. We also thank Matthew Pieri and Satya Gontcho A Gontcho for their comments and feedback through the DESI review process. JF was supported by a Science and Technology Facilities Council (STFC) studentship. AFR acknowledges support by an STFC Ernest Rutherford Fellowship, grant reference ST/N003853/1. AP was supported by the Royal Society. AP and AFR were further supported by STFC Consolidated Grant number ST/R000476/1. This work was partially supported by the UCL Cosmoparticle Initiative. This research is supported by the Director, Office of Science, Office of High Energy Physics of the U.S. Department of Energy under Contract No. DE–AC02–05CH1123, and by the National Energy Research Scientific Computing Center, a DOE Office of Science User Facility under the same contract; additional support for DESI is provided by the U.S. National Science Foundation, Division of Astronomical Sciences under Contract No. AST-0950945 to the National Optical Astronomy Observatory; the Science and Technologies Facilities Council of the United Kingdom; the Gordon and Betty Moore Foundation; the Heising-Simons Foundation; the National Council of Science and Technology of Mexico, and by the DESI Member Institutions. The authors are honored to be permitted to conduct astronomical research on Iolkam Du'ag (Kitt Peak), a mountain with particular significance to the Tohono O'odham Nation.

\bibliographystyle{JHEP.bst}
\bibliography{ms}

\clearpage

\begin{appendix}

\section{The quasar auto-correlation}
\label{sec:Quasar auto-correlation}

\begin{figure}
\centering
\includegraphics[height=0.34\textheight]{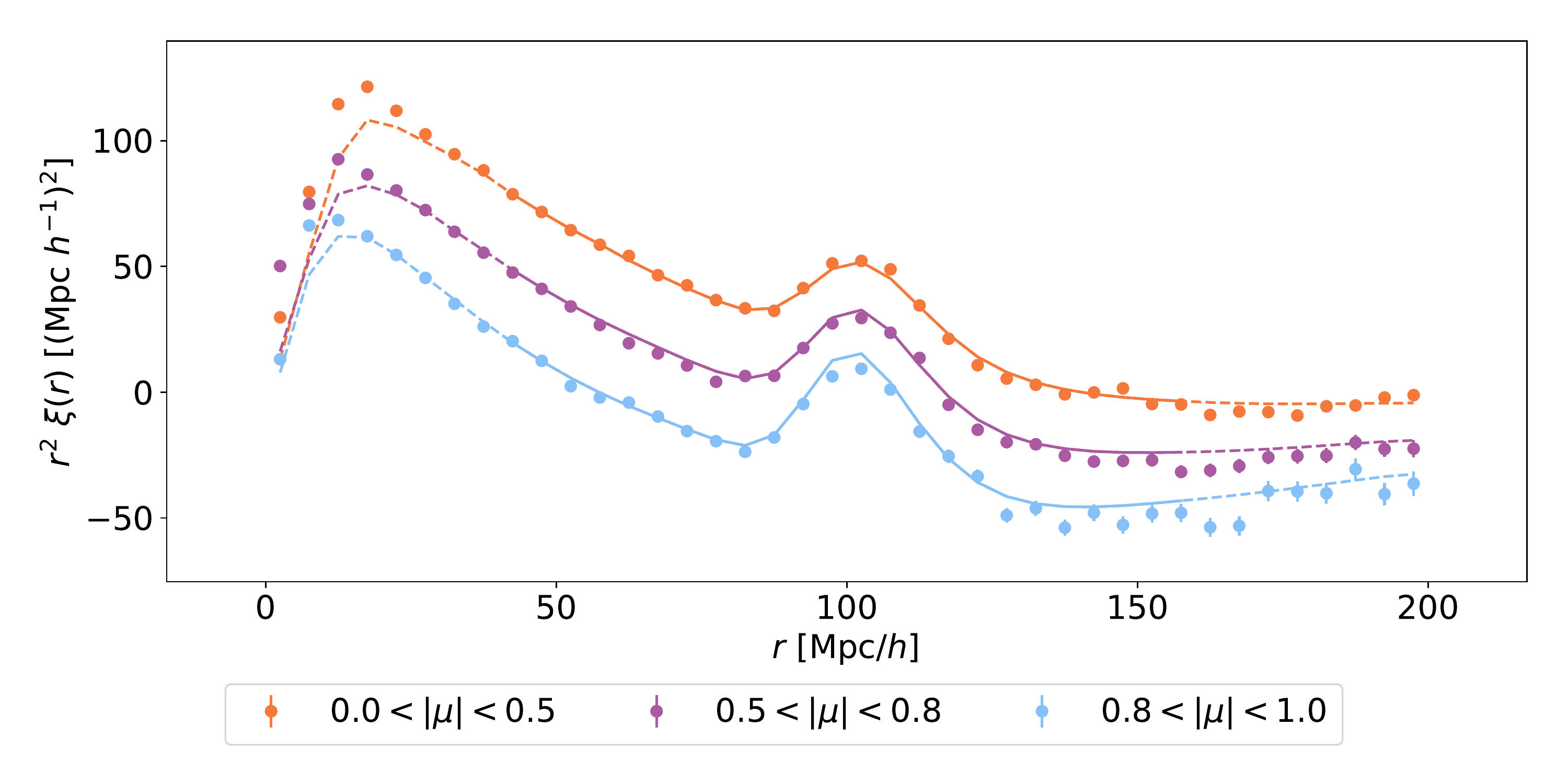}
\caption{The auto-correlation of QSOs, as measured from ten realisations of \texttt{CoLoRe}. The fit is generally good in the fitted region, though the correlation on smaller scales than this is evidently too high.}
\label{fig:corr_plot_QSO_auto}
\end{figure}

We measure the quasar (QSO) auto-correlation on ten QSO catalogues from ten realisations of \texttt{CoLoRe} and combine our results. Correlations are computed as the weighted sum of pairs of QSOs in a grid of parallel and perpendicular separation bins. We divide the sky into \texttt{HEALPix} pixels, computing ``data-data'', ``data-random'' and ``random-random'' correlations in each one using a random catalogue of QSOs. This random catalogue has the same number density distribution of QSOs as that in the mock data, and is generated by \texttt{LyaColoRe}. The different correlation types are then combined using the Landy-Szalay estimator \cite{Landy:1993ApJ...412...64L}, and the covariance is estimated via sub-sampling across \texttt{HEALPix} pixelisations of all 10 realisations (as described in \secref{subsec:Measuring correlation functions}). As in \secref{subsec:Measuring correlation functions}, all correlations are computed using \texttt{picca}\footnote{Publicly available at \href{https://github.com/igmhub/picca}{https://github.com/igmhub/picca}.}. A Kaiser model \cite{Kaiser:1987MNRAS.227....1K} is then fitted to the measurement, leaving free parameters describing the location of the BAO peak and the QSO bias $b_{\delta,\mathrm{QSO}}$. As in \secref{subsec:Fitting the correlation functions}, we also leave free parameters describing the smoothing of the input power spectrum in the parallel and perpendicular directions. As in \secref{subsec:Fitting the correlation functions}, we fit only in the range $40<r\ [\mathrm{Mpc}/h]<160$ as the lognormal approximation begins to break down below this range. The resultant fit is very good in the fitted region, as shown in Figure~\ref{fig:corr_plot_QSO_auto}. We measure a QSO bias of $3.57\pm 0.01$ at an effective redshift of $z=2.20$, consistent with the input value of $3.56$ to within $1\sigma$.

\section{Redshift-space distortions: implementation details}
\label{sec:RSD implementation}

As described in \secref{subsubsec:Adding redshift-space distortions}, adding RSDs to our skewers requires the calculation of a matrix of weights $W_{ij}$ to map each skewer's real-space cells $\tau^x_j$ to redshift-space cells $\tau^s_i$ via the matrix equation $\tau^s_i = W_{ij}\tau^x_j$. $W_{ij}$ is determined by representing each cell as a top-hat function in real space, mapping this profile into redshift space according to the choice of kernel $K$, and calculating the overlap with each redshift-space cell:
\begin{equation}
\label{Weights matrix}
    W_{ij} = \int_{s^l_j}^{s^u_j} P(s-x_i-v_{r,i}|T_i,d_i) \mathrm{d}s,
\end{equation}
where ${s^l_j}$ and ${s^u_j}$ are the lower and upper boundaries of cell $j$ in redshift-space, and $P(x|T,d)$ describes the profile of the real-space cell when mapped into redshift space. $P(x|T,d)$ is dependent on the distance from the centre of the cell $x$, the temperature of the gas $T$ and the half-width of the cell $d$. The form of $P$ is determined by the choice of $K$ (defined in equation \eqref{RSD convolution}):
\begin{equation}
\label{Profile in redshift space}
    P(x|T,d) = \frac{1}{2d}\int_{-d}^{d} K(x-y|T) \mathrm{d}y.
\end{equation}
As such, in the case of $K$ chosen to be a Dirac delta function, the redshift-space cell is represented by a top-hat function (as it was in real space).

In order to account for thermal broadening when adding RSDs to our skewers, we must instead choose our kernel $K$ to be defined by
\begin{equation}
    K(x|T) = \frac{1}{\sqrt{2 \pi} \sigma_v(T)} \exp \bigg( -\frac{x^2}{2\sigma^2_v(T)} \bigg),
\end{equation}
where $\sigma_v(T)$ is the thermal velocity dispersion, which we approximate as in \cite{McDonald:2001ApJ...562...52M} by
\begin{equation}
    \sigma_v(T) = 9.1 \bigg(\frac{T}{10,000\rm{K}}\bigg)^{1/2}\ \mathrm{km\ s}^{-1},
\end{equation}
for temperature $T(z,\textbf{x}) = T_0(z) \rho(z,\textbf{x}) ^{\gamma(z) - 1}$. As described in \secref{sec:Parameter tuning}, for the purposes of this work we fix $\gamma = 1.5$. We also fix $T_0 = 10,000$ K in line with \cite{Slosar:2011JCAP...09..001S} and consistent with literature values~\cite{Ricotti:2000ApJ...534...41R,McDonald:2001ApJ...562...52M,Hiss:2018ApJ...865...42H}. Of course, these values can easily be updated to follow a more complex redshift dependence for any uses of \texttt{LyaCoLoRe} where thermal broadening effects become significant. Evaluating equation \eqref{Profile in redshift space} for this choice of $K$ yields a cell profile in redshift space defined by
\begin{equation}
    P(x|T,d) = \frac{1}{4d}\bigg[\mathrm{erf}\bigg(\frac{x+d}{\sqrt{2}\sigma_v(T)}\bigg) - \mathrm{erf}\bigg(\frac{x-d}{\sqrt{2}\sigma_v(T)}\bigg)\bigg],
\end{equation}
and the matrix of weights can then be computed as per equation \eqref{Weights matrix}.

\section{The \Lya-HCD cross-correlation}
\label{sec:The Lya-HCD cross-correlation}

In Figure~\ref{fig:corr_plot_systematics}, we showed the cross-correlation between \Lya\ absorption and high column density systems (HCDs) from ten realisations of \texttt{LyaCoLoRe}, comparing it to a linear theory model similar to that used to describe the cross-correlation with quasars (QSOs). This model assumes that HCDs have the same clustering as dark matter halos, with a large-scale bias of approximately 2.0. However, in a QSO survey, HCDs are only detected when they are absorbing light from a background QSO, and this observational bias is not taken into account in our modelling. In this appendix, we propose that this bias results in an asymmetry in the measured correlation function. We present a qualitative description of this effect and explain our choice to use only HCDs detected far away from the QSO in Figure~\ref{fig:corr_plot_systematics} in this context.

\begin{figure}
\centering
\includegraphics[height=0.34\textheight]{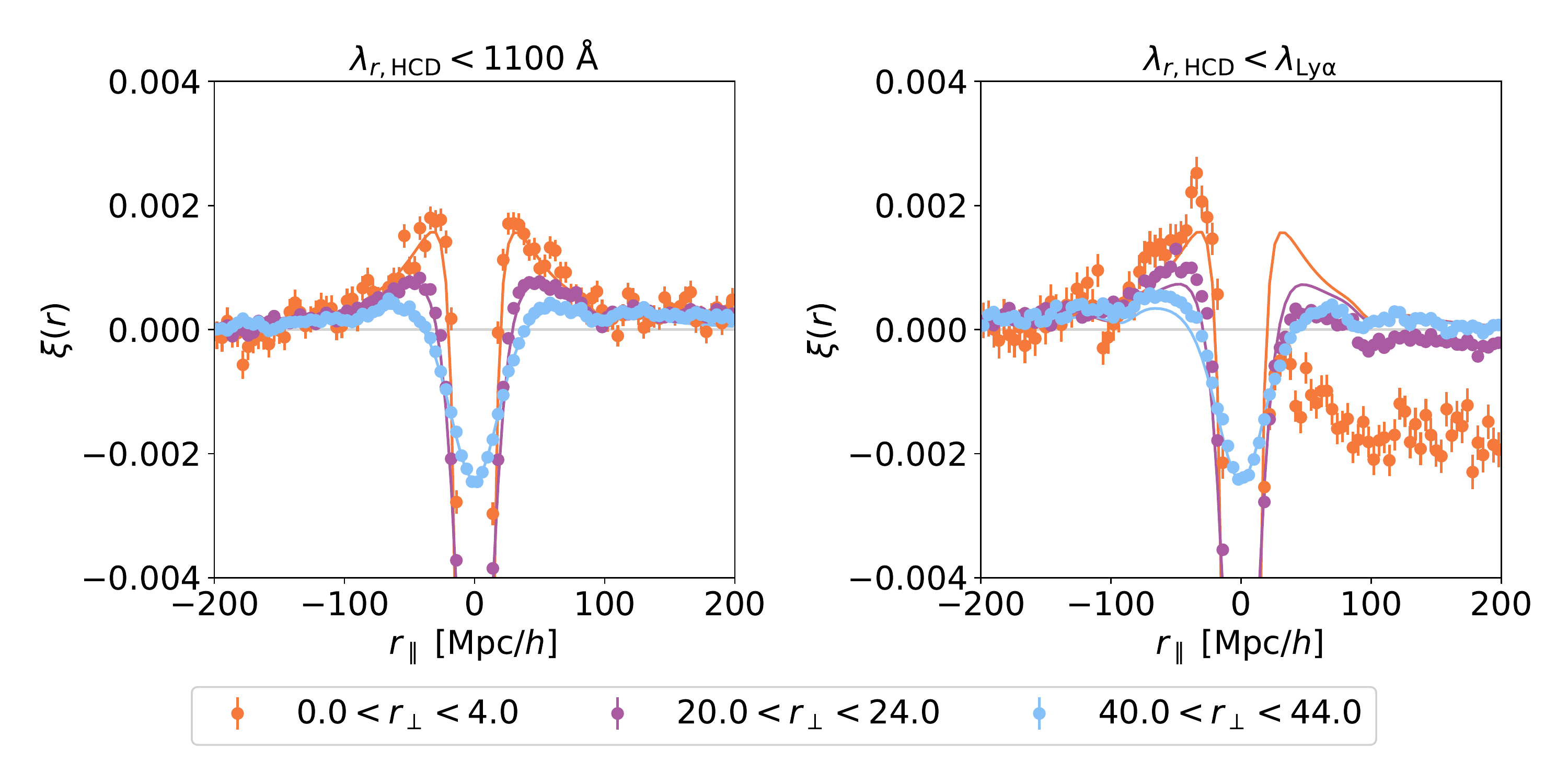}
\caption{The \Lya-HCD cross-correlation, plotted against $r_\parallel$ for different bins of $r_\perp$. The left panel shows the combined measurement from ten realisations using an HCD catalogue that only includes HCDS with rest-frame wavelength less than 1100~\AA\ (as in the right panel of Figure \ref{fig:corr_plot_systematics}). The right panel shows the correlation measured from one realisation when using an HCD catalogue that includes HCDs in the full rest-frame wavelength range, up to $\lambda_{\mathrm{Ly}\alpha}=1215.67$~\AA. The solid lines in both panels show the same fitted correlation as in Figure \ref{fig:corr_plot_systematics}: the joint fit of the \Lya\ auto-correlation and \Lya-HCD cross-correlation from ten realisations of \texttt{LyaCoLoRe}.}
\label{fig:corr_plot_vs_rp_lya_dla_cross}
\end{figure}

In the left panel of Figure~\ref{fig:corr_plot_vs_rp_lya_dla_cross} we show the same measurement of the \Lya-HCD cross-correlation as in the right panel of Figure~\ref{fig:corr_plot_systematics}, this time plotting the correlation against $r_\parallel$ in 3 narrow bins of $r_\perp$. The solid lines show the model obtained by fitting this measurement jointly with the \Lya\ auto-correlation. The model is generally able to fit the measurement well, though some small residuals remain at large $r_\parallel$. These are visible at large separations in the right panel of Figure \ref{fig:corr_plot_systematics}, accentuated due to the factor of $r^2$ in that plot.

In the right panel of Figure~\ref{fig:corr_plot_vs_rp_lya_dla_cross} we plot the \Lya-HCD cross-correlation measured on one realisation of \texttt{LyaCoLoRe}, this time using a full HCD catalogue (with no maximum rest-frame wavelength). The solid lines are the exact same lines as in the left panel. It is clear from this plot that there is a strong asymmetry in the data, and the model used to fit the data in the left panel does not fit this measurement well.

\begin{figure}
\centering
\includegraphics[height=0.2\textheight]{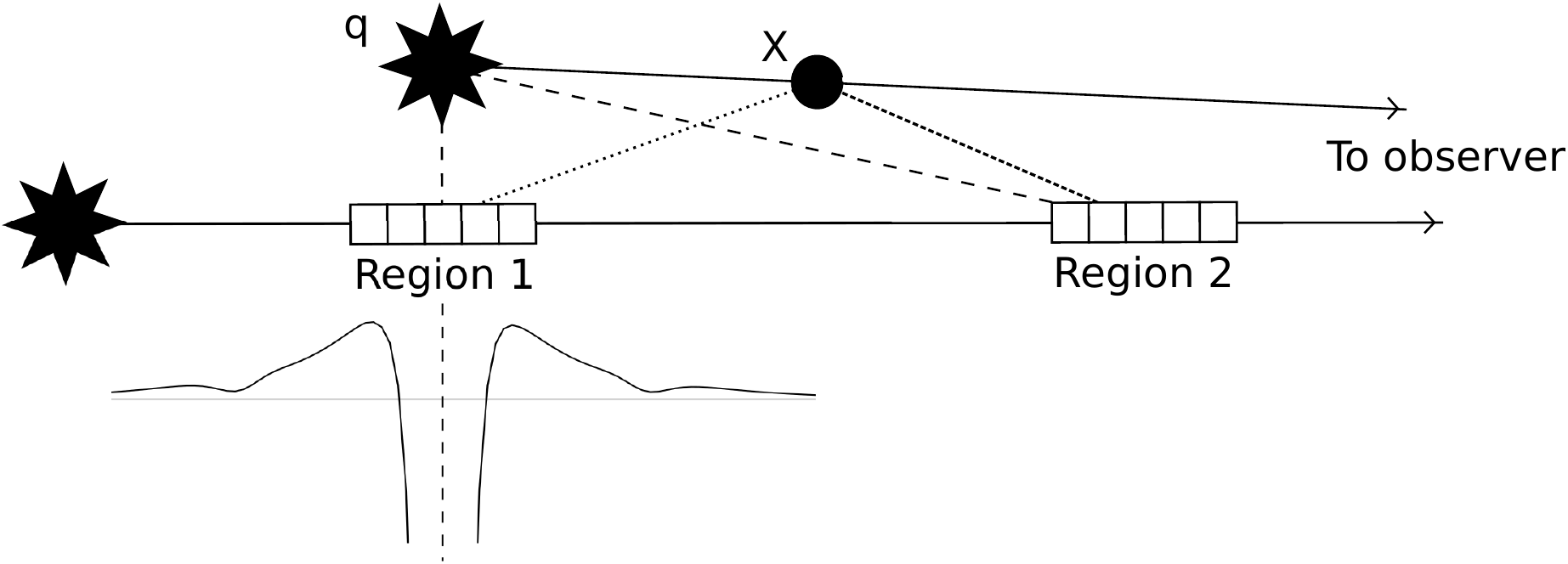}
\caption{A diagram showing the geometry of the setup involved when measuring the \Lya-HCD cross-correlation between two near-parallel skewers. Given the proximity of the QSO $q$ to ``Region 1'' of the lower skewer, we expect to measure biased values of $\delta_F$ for cells corresponding to that region. The cells' values will tend to be reduced or boosted according to the \Lya-QSO cross correlation, as indicated beneath ``Region 1''. This biasing is then imprinted on the correlation between an HCD $X$ and the skewer.}
\label{fig:lya_dla_cross_diagram}
\end{figure}

We propose that this asymmetry is a consequence of the observational bias that is inherently present in our HCD sample, and the dependence on the \Lya-QSO cross-correlation that this induces. According to the density-QSO cross-correlation, a QSO $q$ will tend to have dense regions of gas around it. In relation to an HCD $X$ in $q$'s spectrum, these dense regions will be located at small $r_\perp$ and $r_\parallel \simeq r_{Xq}$, the distance between $X$ and $q$ (as $X$ is constrained to lie directly along the line of sight between $q$ and the observer). This preferential location of dense regions of gas will imprint a feature in the correlation between $X$ and neighbouring skewers of $\delta_F$ at these specific separations. Referring to the diagram in Figure \ref{fig:lya_dla_cross_diagram}, we can see that the cells of $\delta_F$ in Region 1 will tend to be significantly biased according to the \Lya-QSO cross-correlation. Thus, we will see a feature in the correlation between HCD $X$ and its neighbouring skewer corresponding to this region. The shape of this feature is determined by the shape of the \Lya-QSO cross-correlation at small $r_t$, as shown beneath Region 1. The cells in Region 2 will not be significantly affected by the presence of $q$, and so we would not expect to see a feature here.

Summing over HCD-pixel pairs in order to compute the full \Lya-HCD cross-correlation will average out most of the signal, but a small, asymmetric residual will remain, as seen in the right panel of Figure~\ref{fig:corr_plot_vs_rp_lya_dla_cross}. The contribution from each HCD will carry a similar signature but the signature will be centred at different values of $r_\parallel$ due to the different values of $r_{Xq}$ for each $X$-$q$ pair. Certainly though, the sign of $r_{Xq}$ will always be the same as an HCD is always less distant than its host QSO. Using \texttt{picca}'s definition of the sign of $r_\parallel$, this means that $r_{Xq}>0$ for all $X$ and $q$. As a result, we will see a reduction of the \Lya-HCD cross-correlation for all $r_\parallel>0$, due to the strong reduction in $\delta_F$ at the centre of regions such as Region 1 in Figure \ref{fig:lya_dla_cross_diagram}. This is only apparent for $r_\perp$ small as the reduced area shown in Figure~\ref{fig:lya_dla_cross_diagram} is narrow. We also see a secondary effect: a boost in the \Lya-HCD cross-correlation for small, negative $r_\parallel$. This is a result of the small boost in $\delta_F$ on the right-hand side of Region 1 in Figure \ref{fig:lya_dla_cross_diagram}, which appears at $r_\parallel<0$ for HCDs that are very close to their host QSOs. This effect extends to larger values of $r_\perp$ due to the greater width of the boosted area (relative to that which is reduced). 

Whilst interesting, these effects are very small. In order to assess their visibility in current/future studies, we would need to carry out tests using a more realistic mock dataset. This would involve using the entire data reduction pipeline --- including continuum fitting and the use of a distortion matrix --- and is beyond the scope of this work. As an approximate comparison, we observe that the size of the deviation of points in the $0.0<r_\perp<4.0$ bin in the right panel of Figure \ref{fig:corr_plot_vs_rp_lya_dla_cross} is approximately an order of magnitude smaller than the size of the error bars in the uppermost two panels of Figure 2 of \cite{Pérez-Ràfols:2018MNRAS.473.3019P}\footnote{It should be noted that \cite{Pérez-Ràfols:2018MNRAS.473.3019P} includes only HCDs at least 5000~km/s away from their host quasar, equivalent to a rest-frame wavelength cut of approximately 1195~\AA. We choose to use $\lambda_{r,\mathrm{HCD}}<\lambda_{\mathrm{Ly}\alpha}$ in the right panel of Figure \ref{fig:corr_plot_vs_rp_lya_dla_cross} in order to explain the relationship between the geometry of the problem and the observed effect more clearly.}.

Making such an extreme cut in rest-frame wavelength greatly reduces the number of HCDs in our catalogue. In this work we use approximately 30 times the number of skewers as DESI will have, and so this reduction does not cause us any concern. For studies from real surveys, however, maximising the scientific value of their data will be of much greater importance. As such, we would recommend the development of a new model to account for the effects described above using the measured \Lya-QSO cross-correlation. Alternatively, a catalogue of random HCDs, uncorrelated with the \Lya\ forest, could be generated and used to quantify these effects before accounting for them appropriately. Either way, further tests are needed in order to understand more fully the effect described in this Appendix, particularly if new modelling is required for future \Lya-HCD cross-correlation measurements.

\end{appendix}

\end{document}